\documentclass[11pt]{aastex}
\usepackage{emulateapj5}
\usepackage{psfig}
\usepackage{pslatex}

\begin{document}
\bibliographystyle{plain}

\submitted{submitted to ApJ}
\title{Rayleigh-Taylor stability of a strong vertical magnetic
field at the Galactic center confined by a disk threaded with horizontal
magnetic field}

\author{ B. D. G. Chandran\footnote{e-mail: benjamin-chandran@uiowa.edu} \\
\affil{Dept. Physics \& Astronomy, University of Iowa, Iowa City, IA}
}

\begin{abstract}

Observations of narrow radio-emitting filaments near the Galactic
center have been interpreted in previous studies as evidence of a
pervasive vertical (i.e. $\perp$ to the Galactic plane) milliGauss
magnetic field in the central $\sim 150$ pc of the Galaxy. A simple
cylindrically symmetric model for the equilibrium in this central
region is proposed in which horizontal (i.e. $\parallel$ to the
Galactic plane) magnetic fields embedded in an annular band of
partially ionized molecular material of radius $\sim 150$ pc are
wrapped around vertical magnetic fields threading low-density hot
plasma. The central vertical magnetic field, which has a pressure that
significantly exceeds the thermal pressure of the medium, is confined
by the weight of the molecular material. The stability of this
equilibrium is studied indirectly by analyzing a uniformly rotating
cylinder of infinite extent along the $z$ axis in cylindrical
coordinates $(r,\theta,z)$, with low-density plasma and an axial
magnetic field at $r< 150$ pc, high-density plasma and an azimuthal
field at $r> 150$ pc, and a gravitational acceleration
$g^{\ast}\propto r$ directed in the $-\hat{\bf r}$ direction.  Simple
profiles are assumed for the density $\rho$, pressure $p$, and field
strength $B$, with the sound speed and Alfv\'en speeds $\propto r$
within the dense plasma. The density profile and gravity tend to
destabilize the plasma, but the plasma tends to be stabilized by
rotation and magnetic tension---since the interface between the high
and low-density plasmas can not be perturbed without bending either
the horizontal or vertical field. Normal modes proportional to
$e^{(im\theta + i k_z z - i \sigma t)}$ with $k_z=0 $ and $m\neq 0$
are studied. Such modes neither
bend nor compress the axial field at $r<150$ pc but allow compressions
of the dense plasma along the azimuthal magnetic field that enhance
the destabilizing role of gravity.  It is shown analytically that when
$\beta= 8\pi p/B^2$ is small and the dense plasma is supported against
gravity primarily by rotation, the necessary and sufficient condition
for stability to $k_z=0$ modes is $|g| < 2|\Omega| a$, where $g =
g^\ast - \Omega ^2 r$ is the effective gravity, $\Omega$ is the
uniform angular velocity, and $a$ is the sound speed in the dense
plasma.  Since the effective gravity is determined by the degree to
which magnetic (and to a lesser degree pressure) forces support the
dense plasma, the stability criterion gives an upper limit on the
strength of the axial magnetic field, which is $\sim 1 $ mG for
Galactic-center parameters.

\end{abstract}

\section{Introduction}

The central $\sim 150$ pc of the Galaxy host a phenomenon that is
unique within the Galaxy: narrow, ``vertical'' (i.e. $\perp$ to the
Galactic plane) filaments of radio emission that are typically
a few tens of parsecs long and only a fraction of a parsec wide
(Morris
1996). Polarization measurements indicate that the  magnetic
fields in the Radio Arc and Northern
Thread are parallel to these filaments
(Tsuboi et~al.\ 1986, Reich 1994, Tsuboi et~al.\ 1995, Lang
et~al.\ 1999b), and this alignment is presumed to be typical of the
other filaments as well. In virtually all cases the filaments
appear to be in contact with a molecular cloud at some point along
their length. The absence of bending of these filaments by the
randomly moving clouds has been interpreted as evidence of a
milliGauss lower limit to the field strength in the filaments
(Yusef-Zadeh \& Morris 1987a---this point is discussed further in
section \ref{sec:review}). The apparent impossibility of forming or
confining narrow filaments with such tremendous magnetic
pressures---well in excess of the thermal pressure of the
medium---suggests that if a mG vertical field is present, it pervades
the central $\sim 150$ pc of the Galaxy, and the filaments are simply
those flux tubes that possess large populations of relativistic
electrons (Morris \& Serabyn 1996).  

A strong pervasive magnetic field at the Galactic center would be a
natural consequence of radial inflow in the Galactic disk (Sofue \&
Fujimoto 1987, Morris \& Serabyn 1996), and
magnetic-pressure forces would over time cause the volume-filling
magnetic flux tubes to rise buoyantly away from the Galactic mid-plane,
forcing field lines that penetrate the mid-plane into a vertical
orientation (Chandran et~al.\ 2000). It is not clear, however, that
such a scenario is inevitable, as the rate of radial inflow in the
Galactic disk is not well known and since sufficient turbulent
resistivity can inhibit the central concentration of the field (Lubow,
Papaloizou, \& Pringle 1994). Alternative explanations of the
filaments have been explored by a number of authors (Chudnovsky
et~al.\ 1986, Benford 1988, Heyvaerts, Norman, \& Pudritz 1988, Lesch
\& Reich 1992, Rosso \& Pelletier 1993, Shore \& LaRosa 1999).
If a pervasive mG field is present, it would play a dominant role
in the dynamics of the interstellar medium near the Galactic center,
and would suggest that the Galaxy was born with a magnetic
field $\gtrsim 10^{-7}$~G (Chandran et~al 2000---this point is
discussed further in section~\ref{sec:discussion}).

One of the most pressing questions regarding the
pervasive-mG-field hypothesis is whether such a strong
magnetic field could be stably confined. The volume-dominant phase of
the central region is hot plasma at a temperature $T\sim 10^8$~K which
fills an elliptical region $150 \times 270$ pc (FWHM) with a major
axis that is tilted $20^\circ$ with respect to the Galactic plane
(Yamauchi et~al.\ 1990).  The density of this plasma was estimated to
be $0.03- 0.06 \mbox{ cm}^{-3}$ by Yamauchi et~al.\ (1990) and $\sim
0.3-0.4$ by Koyama et~al.\ (1996).  The mass-dominant phase of the
central region is molecular gas with $n > 10^4 \mbox{ cm}^{-3}$, $T
\sim 70$~K, a filling factor in excess of 0.1, and a vertical
thickness of $\sim 30 $ pc (Bally et~al.\ 1988, Morris \& Serabyn
1996).  If the magnetic field strength is 1~mG, then the magnetic
pressure $B^2/8\pi$ is $\sim 4\times 10^{-8} \mbox{ dyne/cm}^2$, which
far exceeds the thermal pressure of both the hot plasma, $p_{\rm hot}
\sim 4 \times 10^{-10}- 4\times 10^{-9}\mbox{ dyne/cm}^2$, and the
molecular material, $p_{\rm cloud} \sim 10^{-10} \mbox{ dyne/cm}^2$.
The thermal pressure of the ambient medium is thus unable to confine
mG fields, whether they are ubiquitous or in the form of
isolated narrow flux tubes.

\begin{figure*}[h]
\centerline{
\psfig{figure=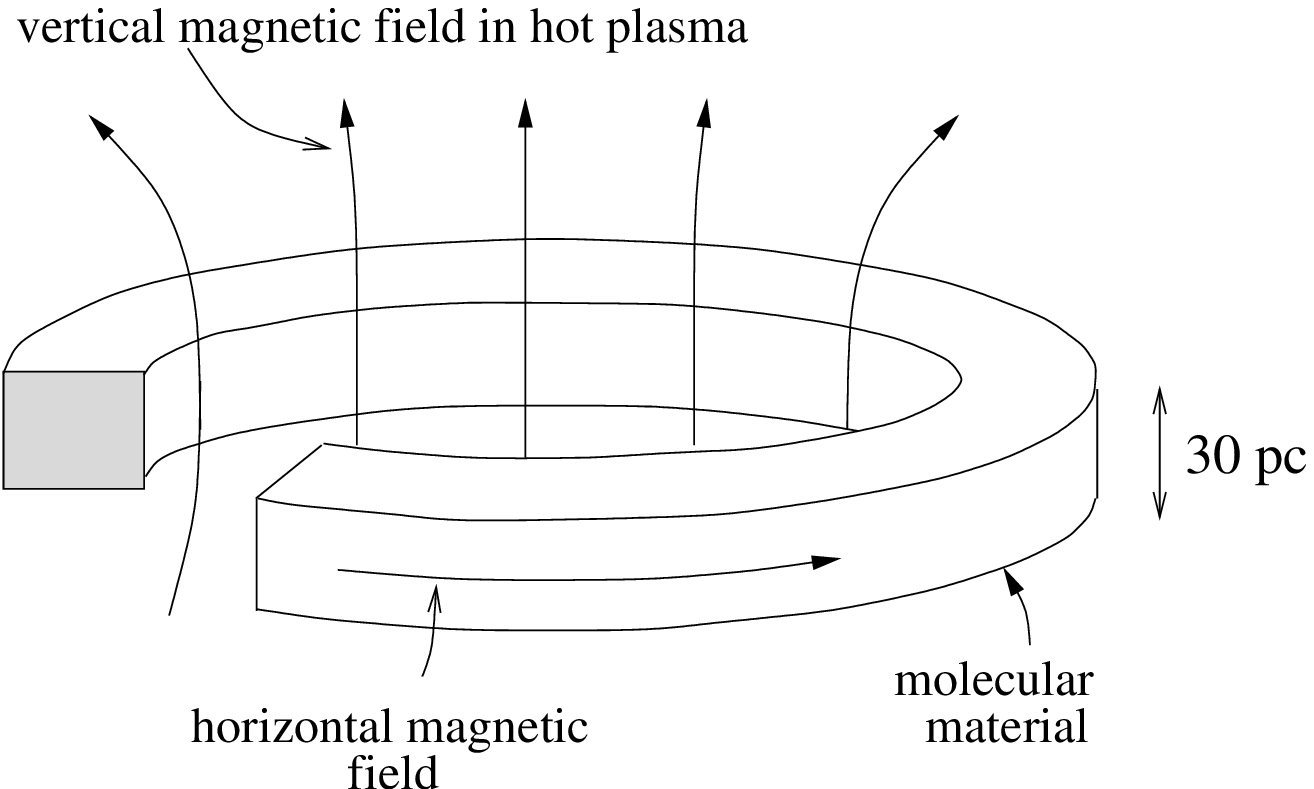,width=4in,clip=}
}
\vspace{2mm}
\caption{A simplified model for an equilibrium of strong
vertical fields at the Galactic center. \label{fig:equil}  }
\end{figure*}

The only force capable of confining a pervasive vertical mG field is
the weight of the molecular material. Interestingly, the polarization
of dust emission indicates that the molecular material is threaded by
horizontal (i.e., $\parallel$ to the Galactic plane) magnetic fields
(Morris \& Serabyn 1996).  Because this horizontal field is
perpendicular to the vertical magnetic field in the hot plasma, it is
difficult for the two types of material to interpenetrate, which
suggests that the weight of the molecular material may be able to
stably prevent a strong vertical field from expanding. Although the
molecular clouds are clumpy, it is expected that the horizontal
magnetic field completely encloses the central region. In this paper,
the equilibrium of figure~\ref{fig:equil} is proposed as
a starting point for investigating the confinement of vertical fields
at the Galactic center.

\begin{figure*}[h]
\centerline{
\psfig{figure=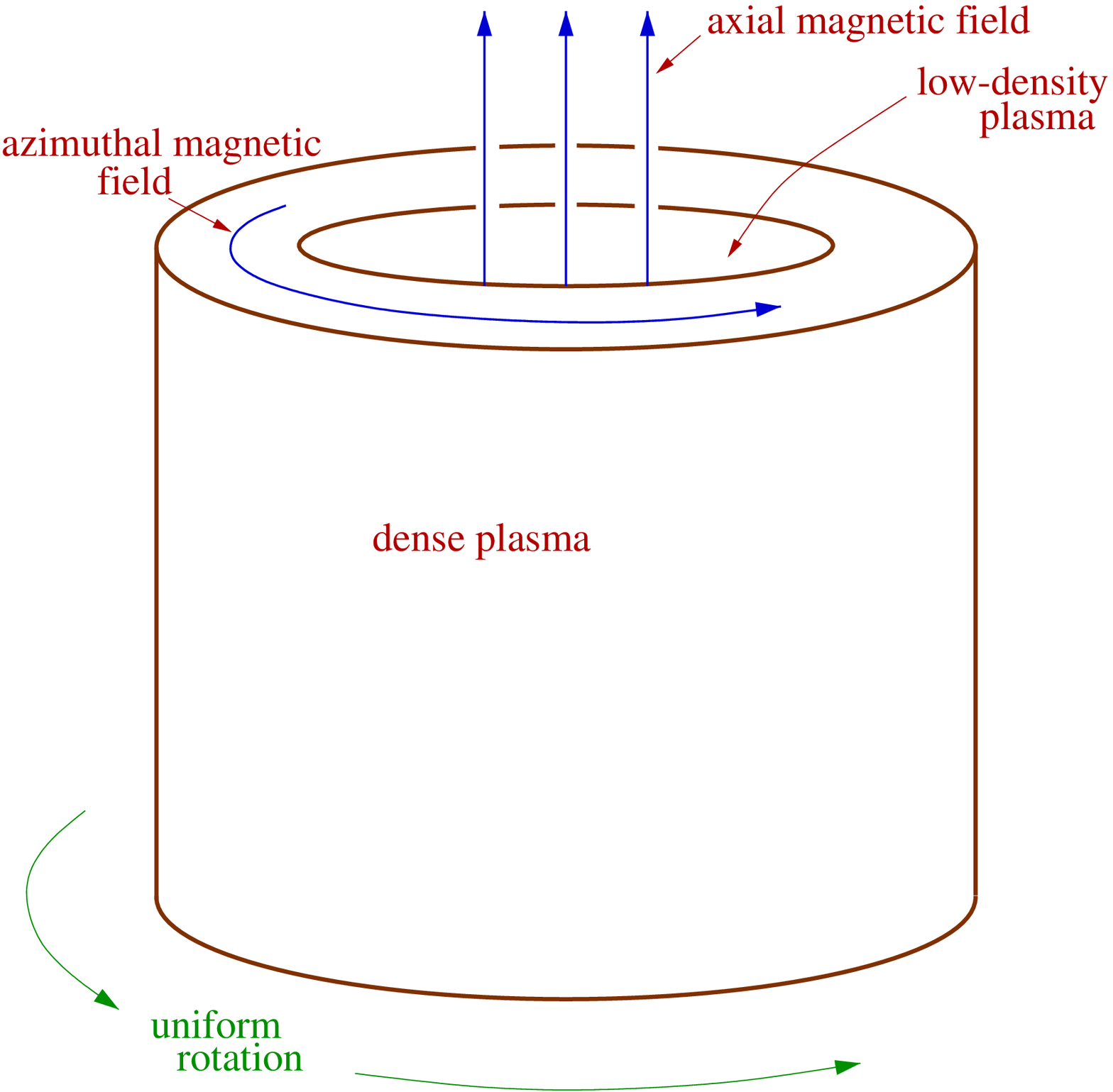,width=4in,clip=}
}
\vspace{2mm}
\caption{The simplified one-dimensional equilibrium that is analyzed in this
paper. \label{fig:equil_picture} }
\end{figure*}

To make a compact analytic treatment possible, the stability of this
equilibrium is explored under a number of additional
simplifying approximations. It is assumed that the fractional
ionization within the molecular clouds is sufficiently large and the
frequencies of any instabilities sufficiently small compared to the
neutral-ion collision frequency that ambipolar diffusion can be
ignored. The partially ionized clouds then behave as a single-species
fluid to which the magnetic field is frozen, or, in other words, as a
dense plasma described by the equations of magnetohydrodynamics (MHD).
The equilibrium is taken to be infinite and invariant along the spin
axis---the $z$ axis in cylindrical coordinates $(r,\theta,z)$--- as
depicted in figure~\ref{fig:equil_picture}.  A constant angular
velocity $\Omega$ is assumed, which removes the magneto-rotational
instability from the problem (Balbus \& Hawley 1998).  An axial
magnetic field and low-density plasma are assumed for $r< 150$ pc, and
an azimuthal magnetic field and dense plasma for $r>150$ pc.  The
equilibrium density $\rho_0$, pressure $p_0$, and field strength $B_0$
are assumed to be proportional to powers of~$r$ within the dense
plasma, and $\beta = 8\pi p_0/B_0^2$ is taken to be constant and
small, which implies that the sound speed~$a$ and Alfv\'en
speed~$v_{\rm A}$ are proportional to~$r$. The gravitational
acceleration~$g^\ast$ is taken to be proportional to~$r$ and to be
directed in the~$-\hat{r}$ direction. If $g^\ast$ arises from stars,
this amounts to assuming a constant mass density of stars throughout
the region occupied by the plasma.  It is also assumed that the plasma
is supported primarily by rotation, so that $v_{\rm A}\ll v_{\rm rot}$
in the dense plasma, where $v_{\rm rot} = \Omega r$.  Although some of
these assumptions are extreme, a number of the key ingredients are
retained: compressibility, the Coriolis force, the suspension of dense
plasma ``above'' low-density plasma, and the orthogonal orientations
of the magnetic fields in the low- and high-density plasmas.

The analysis then focuses on modes which perturb the sharp boundary
between the high- and low-density plasmas. Such perturbations are to
some extent stabilized by magnetic tension, since short-wavelength
disturbances involve significant field-line bending of either the
axial or azimuthal magnetic fields. Uniform rotation also plays a
stabilizing role (e.g., Talwar 1960, Chandrasekhar 1961, Gilman 1970,
Acheson \& Gibbons 1978).  Normal modes of the form ${\bf u} _1 (r)
e^{(ik_z z + im \theta - i \sigma t)}$ with $k_z=0$ and $m\neq 0$ are
studied analtyically. Such modes avoid bending or compressing the
axial magnetic field at $r<150 $ pc and also allow compressions of the
dense plasma along the azimuthal magnetic field that enhance the
destabilizing effects of gravity. It is shown in
sections~\ref{sec:diffeq} through~\ref{sec:bc} that the necessary and
sufficient condition for stability to $k_z=0$ modes is
\begin{equation}
|g| < 2|\Omega| a
\label{eq:stabcrit0} 
\end{equation}
when $a \ll v_{\rm A} \ll \Omega r$, where $g= g^\ast - \Omega^2 r$ is
the effective gravity, and $a$ and $v_{\rm A}$ are evaluated in the
dense plasma.  A physical explanation of 
equation~(\ref{eq:stabcrit0}) is given
in section~\ref{sec:stability}.  Since the effective gravity
is proportional to the partial support of the plasma by the magnetic
field, equation~(\ref{eq:stabcrit0}) can be re-written as an upper
limit on the strength of the vertical field
[equation~(\ref{eq:stabcrit3})]. For Galactic-center parameters, the
upper limit is $\sim 1$~mG. If equation~(\ref{eq:stabcrit0})
is the approximate criterion for the stability of the boundary between
the high and low-density plasmas in systems with differential
rotation, a finite extent along the $z$ axis, and more general radial
profiles---say by evaluating the terms in
equation~(\ref{eq:stabcrit0}) near the inner edge of the dense
plasma---then a pervasive mG vertical magnetic field
can be confined at the Galactic center. Equation~(\ref{eq:stabcrit0})
would then also be relevant to other accreting systems with strong
poloidal magnetic fields.  

Global instabilities with $k_z \neq 0$, local quasi-interchange modes,
and instabilities of the two-dimensional plasma of
figure~\ref{fig:equil} are discussed in section~\ref{sec:local}. In
section~\ref{sec:review}, observational constraints on the
field-strength in the Galactic-center filaments are reviewed.  The
implications of a pervasive mG field at the Galactic center for the
origin of the Galactic magnetic field are discussed in
section~\ref{sec:discussion}.

\section{Differential equation for $k_z=0$ normal modes in the dense
plasma}
\label{sec:diffeq}

\begin{table*}[h]
\begin{center}
\caption{Notation}
\begin{tabular}{lcc}
\hline \hline 
\vspace{-0.2cm} 
\\
Symbol & \hspace{0.3cm} Equation where introduced \hspace{0.3cm} & Meaning \\
\vspace{-0.2cm} 
\\
\hline 
\hline 
\vspace{-0.2cm} 
\\
$\rho$, $p$, ${\bf u}$, ${\bf B}$ & (\ref{eq:cont}), (\ref{eq:mom}) &
mass density, flow velocity, pressure, and magnetic field; (0 \\
& & indicates equilibrium value, 1 indicates perturbed quantity) \\
$g^\ast$, $g$ & (\ref{eq:mom}), (\ref{eq:g})  & true gravity, effective gravity \\
$\Omega$ & (\ref{eq:u0}) & uniform angular velocity \\
$a$, $v_{\rm A}$, $v_{\rm rot}$ & (\ref{eq:a}),
(\ref{eq:va}), (\ref{eq:vrot})  & sound speed, Alfv\'en speed, rotation speed \\
$\gamma$ & (\ref{eq:a})  & adiabatic index \\
$m$, $k_z$, $n$ & (\ref{eq:u1}), (\ref{eq:l3})   & azimuthal, axial, and radial
mode numbers \\
$\sigma$, $\omega$ & (\ref{eq:u1}), (\ref{eq:omega}) & frequency in inertial frame, 
 frequency in rotating frame \\
$h$, $F$, $H$, $\Gamma$, $p_n$ & (\ref{eq:h}),   (\ref{eq:F}),
(\ref{eq:H}), (\ref{eq:Gamma}), (\ref{eq:pn})
& convenient abbreviations \\
$D_{\rm s}$, $D_{\rm A}$, $D_{\rm e}$, $D_{\rm R}$ &
(\ref{eq:ds}), (\ref{eq:da}), (\ref{eq:de}), (\ref{eq:dr}) & dispersion-relation 
polynomials for the sound wave, \\
& & Alfv\'en wave, epicyclic-Alfv\'en mode,
and Rayleigh-Taylor mode \\
$\beta$ & (\ref{eq:beta}) & ratio of thermal to magnetic pressure \\
$x$ & (\ref{eq:x}) & normalized radius \\
$q$ & (\ref{eq:r0s}) & exponent in constant-$\beta$ equilibrium 
($\rho_0\propto r^{-q}$)\\
$\alpha$ & (\ref{eq:gva}), (\ref{eq:c}) &
 order-unity factor in constant-$\beta$ equilibrium\\
$b$ & (\ref{eq:xb}) &  exponent of $r$ in power-law solutions for $u_{1r}$ \\
${\bf \xi}$ & (\ref{eq:xi}) & Lagrangian displacement \\
$\zeta$ & (\ref{eq:zeta})  & measure of the wavenumber \\
$\zeta_{\rm crit}$ & (\ref{eq:zetacrit})  & maximum value of $\zeta$ corresponding to
an unstable mode \\
${\bf F}_g$, ${\bf F}_{\rm C}$, ${\bf F}_p$ & (\ref{eq:fgp}), (\ref{eq:fcp}), 
(\ref{eq:fpfg})  & perturbed gravitational, Coriolis, and pressure forces\\
$B_{\rm vert}$ & (\ref{eq:stabcrit3})
& strength of the pervasive vertical field at the Galactic center\\
\vspace{-0.2cm} 
\\

\hline
\end{tabular}
\end{center}
\end{table*}

The plasma is taken to obey the equations of ideal MHD and
an adiabatic equation of state,
\begin{equation}
\frac{\partial \rho}{\partial t} = - \nabla \cdot (\rho {\bf u}),
\label{eq:cont} 
\end{equation} 
\begin{equation}
\rho\left(\frac{\partial {\bf u} } {\partial t} + {\bf u} \cdot \nabla {\bf u}\right)
= - \nabla\left(p + \frac{B^2}{8\pi}\right) +\frac{1}{4\pi} {\bf B} \cdot \nabla
{\bf B} -\rho g^\ast \hat{r},
\label{eq:mom} 
\end{equation}
\begin{equation}
\frac{\partial {\bf B} }{\partial t} = \nabla \times ({\bf u} \times {\bf B} ),
\mbox{ \hspace{0.3cm} and}
\label{eq:ind} 
\end{equation} 
\begin{equation}
\left(\frac{\partial }{\partial t} + {\bf u} \cdot \nabla\right)\left(
\frac{p}{\rho^\gamma}\right) = 0,
\label{eq:state} 
\end{equation} 
where $\rho$, $p$, ${\bf u}$, and ${\bf B}$ are the density, pressure,
velocity, and magnetic field, $g^\ast$ is the gravitational acceleration,
$(r, \theta, z)$ are cylindrical coordinates, and $\gamma$ is the adiabatic index.
An equilibrium is assumed in which all quantities are functions
of $r$ alone. Equilibrium quantities are denoted with the subscript~0;
\begin{eqnarray} 
\rho_0 & = & \rho_0(r), \\
p_0 & = & p_0(r),\\
{\bf u}_0 & = & \Omega r \, \hat{\theta}, \label{eq:u0}\mbox{ \hspace{0.3cm} and} \\
{\bf B} _0 & = & B_0(r)\, \hat{\theta}. 
\end{eqnarray} 
A uniform angular velocity is assumed,
\begin{equation}
\Omega = \mbox{ constant}, 
\end{equation}
and the effective gravity $g$ is defined by the equation
\begin{equation}
g = g^\ast - \Omega^2 r.
\label{eq:g} 
\end{equation} 
By virtue of equation~(\ref{eq:mom}), 
\begin{equation}
0 = \frac{d}{dr}\left(p_0 + \frac{B_0^2}{8\pi}\right) + \frac{B_0^2}{4 \pi r} + \rho_0 g.
\label{eq:hydrostat} 
\end{equation} 
The sound and Alfv\'en speeds are defined, respectively, by the equations
\begin{equation}
a^2 = \frac{\gamma p_0}{\rho_0}, 
\label{eq:a} 
\end{equation}
and
\begin{equation}
v_{\rm A}^2 = \frac{B_0^2}{4\pi\rho_0}.
\label{eq:va} 
\end{equation} 

Small-amplitude disturbances to the equilibrium, denoted
by the subscript~1, can be decomposed into normal modes of
the form
\begin{equation}
{\bf u}_1 = {\bf u}_1(r) e^{im\theta + ik_z z -i\sigma t},
\label{eq:u1} 
\end{equation}
with analogous equations for ${\bf B} _1$, $\rho_1$, and $p_1$.
The equations governing these disturbances are obtained in the
standard way, by substituting
${\bf u} = {\bf u}_0 + {\bf u}_1$, ${\bf B} = {\bf B}_0 + {\bf B}_1$, etc
into equations~(\ref{eq:cont}) through (\ref{eq:state}) and discarding all
terms containing the product of more than one of the perturbed quantities.
The analysis is confined to the case
\begin{equation}
k_z = 0.
\end{equation}
The frequency $\omega$ of the disturbance in the frame rotating with the plasma is
given by 
\begin{equation}
\omega = \sigma - m \Omega.
\label{eq:omega} 
\end{equation}
From equations~(\ref{eq:cont}) and (\ref{eq:state}), the $\hat{r}$ and $\hat{\theta}$ components
of equation~(\ref{eq:ind}), and the $\hat{\theta}$ component of equation~(\ref{eq:mom}),
one obtains
\begin{equation}
B_{1r} = - \frac{m B_0 u_{1r}}{\omega r} ,
\label{eq:b1r} 
\end{equation}
\begin{equation} 
B_{1\theta} = - \frac{i}{\omega} \frac{d}{dr}(u_{1r}B_0),
\label{eq:b1t} 
\end{equation}
\begin{equation}
u_{1\theta} = - \frac{i}{D_{\rm s}} \left(
hu_{1r} + \frac{m a^2}{r} \frac{ d u_{1r}}{dr} \right),
\label{eq:u1t} 
\end{equation}
\begin{equation}
\rho_1 = \frac{-iu_{1r}}{\omega}
\left[ \frac{\rho_0}{r} \left( 1 + \frac{mh}{D_{\rm s}}\right) 
+ \frac{d\rho_0}{dr}\right] 
- \frac{i\omega \rho_0}{D_{\rm s}} \frac{ d u_{1r}}{dr}, 
\label{eq:rho1} 
\end{equation}
and
\begin{equation}
p_1 = \frac{-iu_{1r}}{\omega}
\left[ \frac{a^2 \rho_0}{r} \left( 1 + \frac{mh}{D_{\rm s}}\right) 
+ \frac{dp_0}{dr}\right] 
- \frac{i\omega \rho_0 a^2}{D_{\rm s}} \frac{ d u_{1r}}{dr},
\label{eq:p1} 
\end{equation}
where 
\begin{equation}
h = 2\Omega \omega - \frac{m}{r} \left( g - \frac{a^2}{r}\right), 
\label{eq:h} 
\end{equation}
and
\begin{equation}
D_{\rm s} = \omega^2 - \frac{m^2 a^2}{r^2}.
\label{eq:ds} 
\end{equation} 
Upon substituting equations~(\ref{eq:b1r}) through (\ref{eq:p1})
into the $\bf \hat{r}$ component of equation~(\ref{eq:mom}),
one obtains
\begin{equation}
\frac{d}{dr} \left[
\rho_0 \left( v_{\rm A}^2 + \frac{a^2 \omega^2}{D_{\rm s}}\right) 
r \frac{du_{1r}}{dr}\right] + rFu_{1r} = 0,
\label{eq:diffeq} 
\end{equation}
where
\[
F = \rho_0 D_{\rm A} - \frac{\rho_0 h^2}{D_{\rm s}} +
\frac{1}{r}\frac{d}{dr}\left( \frac{\rho_0 a^2  m h}{D_{\rm s}}\right)
\]
\begin{equation}
+ \frac{d}{dr}\left(\frac{\rho_0 a^2}{r}\right)
+ \frac{B_0^2}{4\pi r^2} 
 - \rho_0 r \frac{d}{dr}\left(\frac{g}{r}\right),
\label{eq:F} 
\end{equation}
and where 
\begin{equation}
D_{\rm A} = \omega^2 - \frac{m^2 v_{\rm A}^2}{r^2}.
\label{eq:da} 
\end{equation} 
For isothermal equilibria and $\gamma = 1$, equation~(\ref{eq:diffeq}) 
reduces to the $k_z = 0 $ limit of 
equation (2.13) of Acheson \& Gibbons (1978).

Because $k_z =0$, 
$u_{1z}$ and $B_{1z}$ decouple from the other variables and exhibit
purely stable behavior: the $\hat{z}$ component
of equation~(\ref{eq:ind})  gives
\begin{equation}
B_{1z} = - \frac{m B_0 u_{1z}}{\omega r},
\label{eq:b1z} 
\end{equation}
and the $\hat{z}$ component of equation~(\ref{eq:mom}) then gives
\begin{equation}
D_{\rm A} u_{1z} = 0.
\label{eq:u1z} 
\end{equation}

\section{Differential equation for $k_z=0$ normal modes in the 
constant-$\beta$ equilibrium}
\label{sec:eq} 

It is now assumed that the equilibrium has
constant $\beta$, where
\begin{equation} 
\beta = \frac{ 8 \pi p_0}{B_0^2},
\label{eq:beta} 
\end{equation} 
and that the equilibrium quantities in the dense plasma are proportional to
powers of $r$:
\begin{equation}
{\bf B}_0(r)  = \overline{ B} x^{-s}\hat{\theta},
\label{eq:b0s} 
\end{equation} 
\begin{equation}
p_0(r) = \frac{\beta \overline{ B}^2 x^{-2s}}{8 \pi}, \mbox{ \hspace{0.3cm} and}
\end{equation} 
\begin{equation}
\rho_0(r) = \overline{ \rho} x^{-q},
\label{eq:r0s} 
\end{equation}
where 
\begin{equation}
x \equiv \frac{r}{r_1},
\label{eq:x} 
\end{equation} 
and $r_1$, $\overline{ B}$, and $\overline{ \rho}$, are
constants. It is also assumed that 
\begin{equation}
g = \overline{ g}x, 
\label{eq:gs0} 
\end{equation}
where $\overline{ g}$ is a constant. Substituting equations~(\ref{eq:b0s})
through (\ref{eq:gs0}) into equation~(\ref{eq:hydrostat}), one finds
that
\begin{equation}
q = 2s + 2,
\label{eq:qs} 
\end{equation} 
\begin{equation}
v_{\rm A} \propto r,
\end{equation}
\begin{equation}
a \propto r,
\end{equation} 
and
\begin{equation}
\frac{g}{r} = \frac{\alpha v_{\rm A}^2}{r^2},
\label{eq:gva} 
\end{equation}
where
\begin{equation}
\alpha = [ (q-2)(1 + \beta) - 2]/2.
\label{eq:c} 
\end{equation} 
It is assumed that $\alpha$ and $q$ are of
order unity. 

Using equations~(\ref{eq:b0s}) through (\ref{eq:c}) in equation~(\ref{eq:F}), one
finds that 
\begin{equation}
F = \overline{ \rho} H x^{-q},
\label{eq:F2} 
\end{equation}
where 
\begin{equation}
H = D_{\rm A} - \frac{h^2}{D_{\rm s}} + (2-q)\frac{a^2}{r^2} \left(1 + \frac{mh}
{D_{\rm s}}\right) + \frac{v_{\rm A}^2 - a^2}{r^2}.
\label{eq:H} 
\end{equation}
Since $D_{\rm A}$, $D_{\rm s}$, $h$, and $H$ are constant within
the dense plasma,
equation~(\ref{eq:diffeq}) has simple power-law solutions,
\begin{equation}
u_{1r} = x^b,
\label{eq:xb} 
\end{equation}
where
\begin{equation}
b^2 + b(2-q) + \Gamma = 0,
\label{eq:quadratic} 
\end{equation}
and 
\begin{equation}
\Gamma = H \left(\frac{v_{\rm A}^2}{r^2} 
+ \frac{a^2 \omega^2}{r^2 D_{\rm s}}\right)^{-1}.
\label{eq:Gamma} 
\end{equation} 

\section{Dispersion relation for $k_z=0$ normal modes
in the constant-$\beta$ equilibrium}
\label{sec:bc} 

The dense plasma is assumed to have an inner edge at 
$r=r_1$, inside of which there is perfectly conducting
plasma of vanishingly small  density:
\begin{equation}
\rho \rightarrow 0 \hspace{0.3cm} \{ \mbox{for $r< r_1$}\}.
\label{eq:rhoaxial0} 
\end{equation} 
Three different cases are considered for the behavior at large~$r$:
(1) the dense plasma has an outer edge at $r=r_2$, outside of which is
non-magnetized plasma of vanishingly small density, (2) there is an
outer edge at $r=r_2$, outside of which is low-density plasma with an
axial magnetic field, and (3) the constant-$\beta$ equilibrium profile
is taken to extend to infinitely large $r$. All three cases lead to
the essentially the same dispersion relation and stability criterion.
In this section, cases (1) and (2) are treated; case
(3) is discussed in appendix~\ref{ap:inf}.

For uniform rotation, the displacement $\bf \xi$ of a fluid element
relative to where it would have been in the equilibrium flow
is given by 
\begin{equation}
{\bf \xi} = \frac{i {\bf u}}{\omega}.
\label{eq:xi} 
\end{equation}
The boundary conditions at each of the boundaries between the 
high and low density plasmas are
\begin{equation}
|| \hat{n} \cdot {\bf \xi} || = 0,
\end{equation}
and
\begin{equation}
\left|\left | p + \frac{B^2}{8 \pi} \right|\right| = 0,
\label{eq:pressb} 
\end{equation} 
where $\hat {n}$ is the unit normal to the perturbed boundary, and
$|| f ||$ denotes the jump in $f$ across the perturbed boundary.
The Lagrangian perturbation of a quantity $f$, denoted
$\delta f$, is related to the Eulerian perturbation, $f_1$, by the equation
$\delta f = f_1 + {\bf \xi} \cdot \nabla f_0$.  
Equation~(\ref{eq:pressb}) indicates
that the Lagrangian total-pressure perturbation 
(magnetic plus thermal) is constant across the sharp
boundary.

Since
\begin{equation}
\rho \rightarrow 0 \hspace{0.3cm} \{ \mbox{for $r< r_1$ or $r> r_2$}\},
\label{eq:rhoaxial} 
\end{equation} 
$a$ and/or $v_{\rm A}$ must be arbitrarily large in the low-density
plasma in order to maintain pressure balance in the equilibrium state.
It is assumed that
$\omega$ is fixed by the dynamics inside the dense plasma, so that
high-frequency Alfv\'en and sound waves in the
low-density plasma are not treated. For the comparatively
low-frequency modes that are considered, the low-density plasma
instantly rearranges itself to avoid the local compressions that
would lead to thermal or magnetic-pressure perturbations.
Within the low-density plasma, the Lagrangian total-pressure
perturbation thus vanishes, as shown in appendix~\ref{ap:axialfield}.
Equation~(\ref{eq:pressb}) then implies that the Lagrangian
total-pressure perturbation in the dense plasma also vanishes at the
sharp boundaries:
\begin{equation}
p_1 + {\bf \xi}\cdot \nabla p_0 + \frac{B_0 B_{1\theta}}{4\pi}
+ \xi \cdot \nabla \frac{B_0^2}{8\pi} = 0 \hspace{0.3cm} \{\mbox{for $r=r_1$ or $r= r_2$}\}.
\label{eq:lagpr} 
\end{equation}

Using equations~(\ref{eq:b1t}) and  (\ref{eq:p1}), one can rewrite
equation~(\ref{eq:lagpr}) as
\begin{equation}
A_1 u_{1r} + A_2 \left(x \frac{du_{1r}}{dx}\right) = 0 \hspace{0.3cm}
\left\{\mbox{for $x=1$ or $\displaystyle x= \frac{r_2}{r_1}$}\right\},
\label{eq:bc} 
\end{equation}
where
\begin{equation}
A_1 = \frac{a^2}{r^2} ( D_{\rm s} + mh),
\label{eq:a1} 
\end{equation}
and
\begin{equation}
A_2 = \frac{\omega^2 a^2 + D_{\rm s} v_{\rm A}^2}{r^2}.
\label{eq:a2} 
\end{equation}
The general solution to equation~(\ref{eq:diffeq}) for the constant-$\beta$
equilibrium is a linear combination of the solutions given in equation~(\ref{eq:xb}),
\begin{equation}
u_{1r} = c_1 x^{b_1} + c_2 x^{b_2},
\label{eq:c1c2} 
\end{equation}
where
\begin{equation}
b_1 = \left( \frac{q}{2} -1\right) +
\sqrt{\left(\frac{q}{2} -1 \right)^2 - \Gamma},
\label{eq:b1b} 
\end{equation} 
and
\begin{equation}
b_2 = \left( \frac{q}{2} -1\right) -
\sqrt{\left(\frac{q}{2} -1 \right)^2 - \Gamma}.
\label{eq:b2b} 
\end{equation} 
Applying equation~(\ref{eq:bc}) at $x=1$ and at
\begin{equation}
x_2 \equiv \frac{r_2}{r_1}
\label{eq:x2} 
\end{equation}
yields the matrix equation
\begin{equation}
\left(\begin{array}{cc}
A_1 + A_2 b_1 & A_1 + A_2 b_2 \\
A_1 + A_2 b_1 & x_2^{b_2 - b_1} [A_1 + A_2 b_2] 
\end{array}
\right)
\left(\begin{array}{c}
c_1 \\
c_2 
\end{array}
\right) = 
\left(\begin{array}{c}
0 \\
0 
\end{array}
\right).
\label{eq:meq} 
\end{equation} 
Non-trivial solutions require that
\begin{equation}
(A_1 + A_2b_1) (A_1 + A_2 b_2) (x_2^{b_2 - b_1} - 1) = 0.
\label{eq:det} 
\end{equation}

One type of normal mode results from setting 
$x_2 ^{b_2 - b_1} = 1$, or, equivalently, 
\begin{equation}
b_2 - b_1 = \frac{2\pi n i}{\ln (r_2/r_1)}, \hspace{0.3cm} n= \pm 1, \pm 2, \dots
\label{eq:l3} 
\end{equation}
Equations~(\ref{eq:b1b}), (\ref{eq:b2b}) and~(\ref{eq:l3}) yield
the dispersion relation
\begin{equation}
\Gamma - p_n = 0,
\label{eq:ldisp1} 
\end{equation}
where
\begin{equation}
p_n = \left[\frac{ \pi n}{\ln(r_2/r_1)}\right]^2 + \left(\frac{q}{2} - 1\right)^2.
\label{eq:pn} 
\end{equation}
The Lagrangian total-pressure perturbation of these modes 
vanishes at $r=r_1$, $r=r_2$, and at $n-1$ discrete
values of $r$ between $r_1$ and $r_2$.  Note that $n=0$ in
equation~(\ref{eq:l3}) does not correspond to an eigenmode, since it
implies $b_1 = b_2$, $c_1 = - c_2$ from equation~(\ref{eq:meq}), and
thus $u_{1r} = 0$.

A second type of normal mode results from having either
\begin{eqnarray} 
A_1 + A_2 b_1 & = & 0 \mbox{ \hspace{0.3cm} and}
\label{eq:op1} \\
c_2 & =&  0, \label{eq:op2} 
\end{eqnarray}
or 
\begin{eqnarray} 
A_1 + A_2 b_2 & = & 0 \mbox{ \hspace{0.3cm} and}
\label{eq:op3} \\
c_1 & =&  0. \label{eq:op4} 
\end{eqnarray}
In either case, one sets $b= - A_1/A_2$ and 
substitutes into equation~(\ref{eq:quadratic}) to obtain
\[
0= \frac{a^4}{r^4} (D_{\rm s} + mh)^2 + 
\]
\begin{equation} 
\left( \frac{v_{\rm A}^2 D_{\rm s}}{r^2}
+ \frac{a^2\omega^2}{r^2}\right)\left[ D_{\rm A}D_{\rm s} - h^2 +
\frac{D_{\rm s}(v_{\rm A}^2 - a^2)}{r^2}\right].
\label{eq:gdisp1} 
\end{equation}
The Lagrangian total-pressure perturbation of these modes vanishes
everywhere---they are the gravest modes of the
system for any given $m$. The right-hand side of equation~(\ref{eq:gdisp1}) is
proportional to $D_{\rm s}$, but the $D_{\rm s}=0$ roots are spurious;
in deriving equation~(\ref{eq:diffeq}) it was assumed that $D_{\rm s}
\neq 0$, and it can be verified that $D_{\rm s} = 0$ does not lead to
normal-mode solutions satisfying the boundary conditions.

Equations~(\ref{eq:ldisp1}) and
(\ref{eq:gdisp1}) can be rewritten in the form
\begin{equation}
0 = D_{\rm e} D_{\rm R} + S,
\label{eq:gform} 
\end{equation}
where
\begin{equation}
D_{\rm e} = \omega^2 - G,
\label{eq:de} 
\end{equation}
\begin{equation}
D_{\rm R} = \omega^2 - \omega \frac{4mg\Omega}{r G }
+ \frac{m^2 }{G} \left( \frac{g^2}{r^2} - \frac{\zeta a^2 v_{\rm A}^2}{r^4}
\right),
\label{eq:dr} 
\end{equation}
\begin{equation}
G = 4\Omega^2 + \frac{\zeta v_{\rm A}^2 }{r^2},
\label{eq:G} 
\end{equation}
and
\begin{equation}
\zeta = \left\{ \begin{array}{ll}
p_n + m^2 -1 & \mbox{ for the roots of equation~(\ref{eq:ldisp1})  }\\
m^2 -1 & \mbox{ for the roots of equation~(\ref{eq:gdisp1}) } 
\end{array} \right. .
\label{eq:zeta} 
\end{equation}
The value of $S$ is different for equation~(\ref{eq:ldisp1}) 
and equation~(\ref{eq:gdisp1}). For equation~(\ref{eq:ldisp1})
\[
S = \frac{4mg\Omega \omega^3}{r G} - \frac{m^2 \omega^2}{G}
\left( \frac{g^2}{r^2} - \frac{\zeta a^2 v_{\rm A}^2}{r^4}\right)
\]
\begin{equation} 
- \frac{a^2}{r^2}\left[
\omega^2 ( q + \zeta) + mq\left( 2\omega \Omega - \frac{mg}{r}\right)
\right],
\label{eq:rl} 
\end{equation}
whereas for equation~(\ref{eq:gdisp1})
\[
S = - \frac{a^2  v_{\rm A}^2}{r^2 (a^2 + v_{\rm A}^2)}
\left[\omega^2\left(1-\frac{m^2 a^2}{v_{\rm A}^2}\right)
+ \frac{m^2(m^2 -1)a^2}{r^2} \right.
\]
\begin{equation} 
\left. + 2m\left( 2\omega\Omega - \frac{mg}{r}\right)\right]
+ \frac{4mg\Omega \omega^3}{r G} - 
\frac{m^2 \omega^2(4\Omega^2 a^2 + g^2)}{G r^2}.
\label{eq:rg} 
\end{equation} 

\begin{figure*}[h]
\vspace{18cm}
\includegraphics{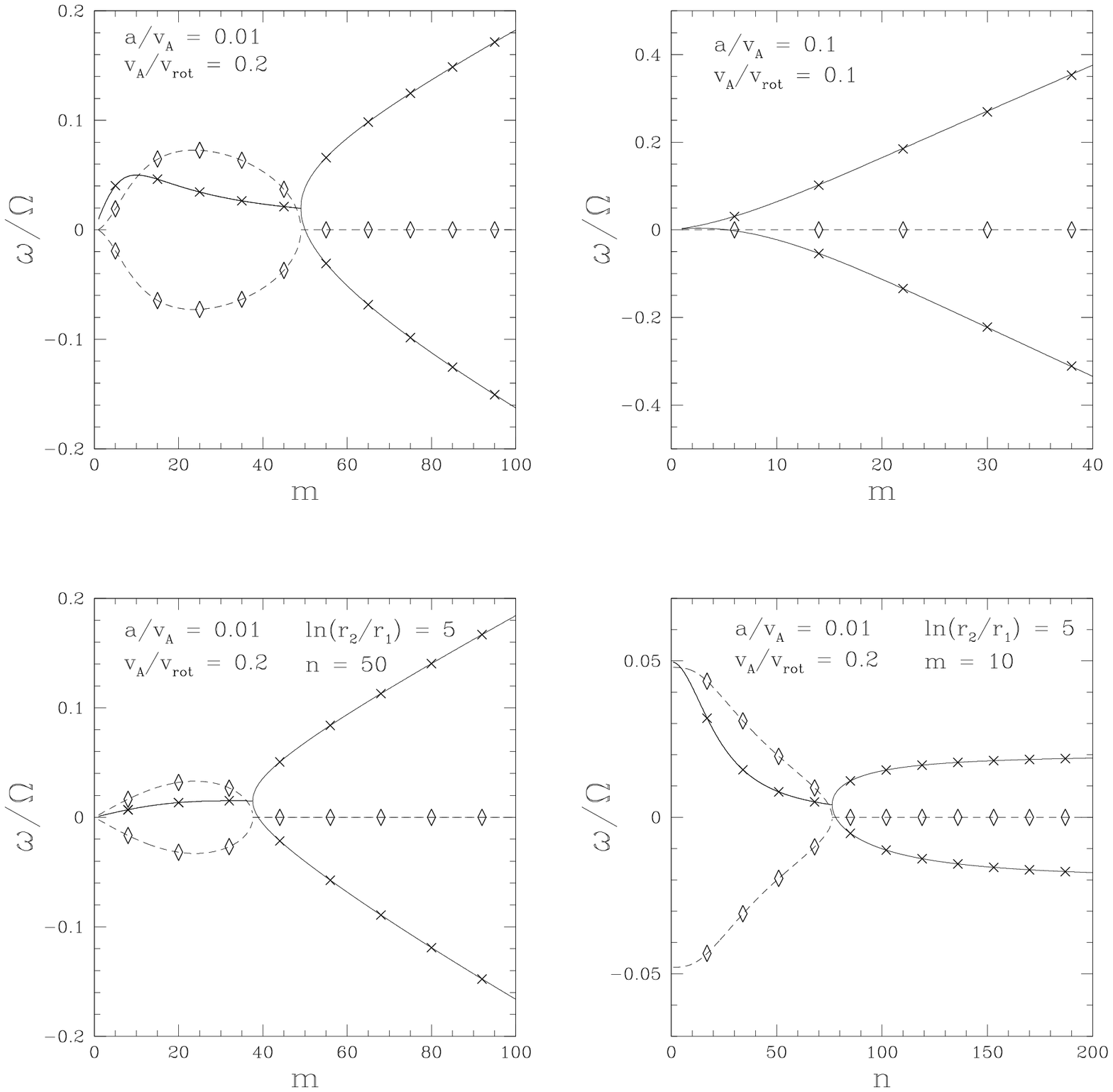}
\caption{Comparison of approximate analytic solution to dispersion
relation for Rayleigh-Taylor modes [equation~(\ref{eq:gs})] and
numerical solutions of equations~(\ref{eq:ldisp1}) and
(\ref{eq:gdisp1}).  The solid (dashed) lines represent the real
(imaginary) part of $\omega/\Omega$ from equation~(\ref{eq:gs}), while
the crosses (diamonds) give the real (imaginary) part of numerical
solutions of equations~(\ref{eq:ldisp1}) and (\ref{eq:gdisp1}). The
upper two panels correspond to equation~(\ref{eq:gdisp1}), and the
lower two panels correspond to equation~(\ref{eq:ldisp1}).  In the
upper-right panel, $a/v_{\rm A}= 0.1$ and $v_{\rm A}/v_{\rm rot} = 0.1$, and the
plasma is stable. In the other three panels, $a/v_{\rm A}= 0.01$ and $v_{\rm A}/v_{\rm
rot} = 0.2$, and the plasma is unstable. For all of the plots,
$q=5$. \label{fig:comparison} }
\end{figure*}

It is now assumed that the dense plasma satisfies the low-$\beta$
condition
\begin{equation}
a \ll v_{\rm A},
\label{eq:lowbeta} 
\end{equation} 
and is supported against gravity primarily by rotation, which implies
that 
\begin{equation}
v_{\rm A} \ll v_{\rm rot},
\label{eq:rotsup} 
\end{equation} 
where
\begin{equation}
v_{\rm rot} \equiv |\Omega r|.
\label{eq:vrot} 
\end{equation}
Equations~(\ref{eq:lowbeta}) and (\ref{eq:rotsup}) imply that
$S$ is negligible compared to the dominant terms in $D_{\rm e} D_{\rm R}$
for all values of $\omega$, $m$, and $p_n$. The approximate
roots of equation~(\ref{eq:gform}) are then obtained by setting
$D_{\rm e} D_{\rm R} = 0.$ 
When $D_{\rm e}=0$,
\begin{equation}
\omega \simeq \pm \sqrt{4\Omega^2 + \zeta v_{\rm A}^2/r^2}.
\label{eq:ae} 
\end{equation}
When $D_{\rm R}=0$,
\begin{equation}
\omega \simeq \frac{ 2 m g \Omega r^2 \pm mv_{\rm A}
\sqrt{\zeta(4\Omega^2 a^2 r^2 - g^2 r^2 + \zeta a^2 v_{\rm A}^2)}}
{4\Omega^2 r^3 + \zeta r v_{\rm A}^2 }.
\label{eq:gs} 
\end{equation}
The errors in equations~(\ref{eq:ae}) and (\ref{eq:gs}) vanish as
$(a/v_{\rm A}) \rightarrow 0$ and $(v_{\rm A}/v_{\rm rot})\rightarrow
0$.  Comparisons between equation~(\ref{eq:gs}) and
numerical solutions of equations~(\ref{eq:ldisp1}) and
(\ref{eq:gdisp1}) are given in figure~\ref{fig:comparison}.

\begin{figure*}[h]
\centerline{
\psfig{figure=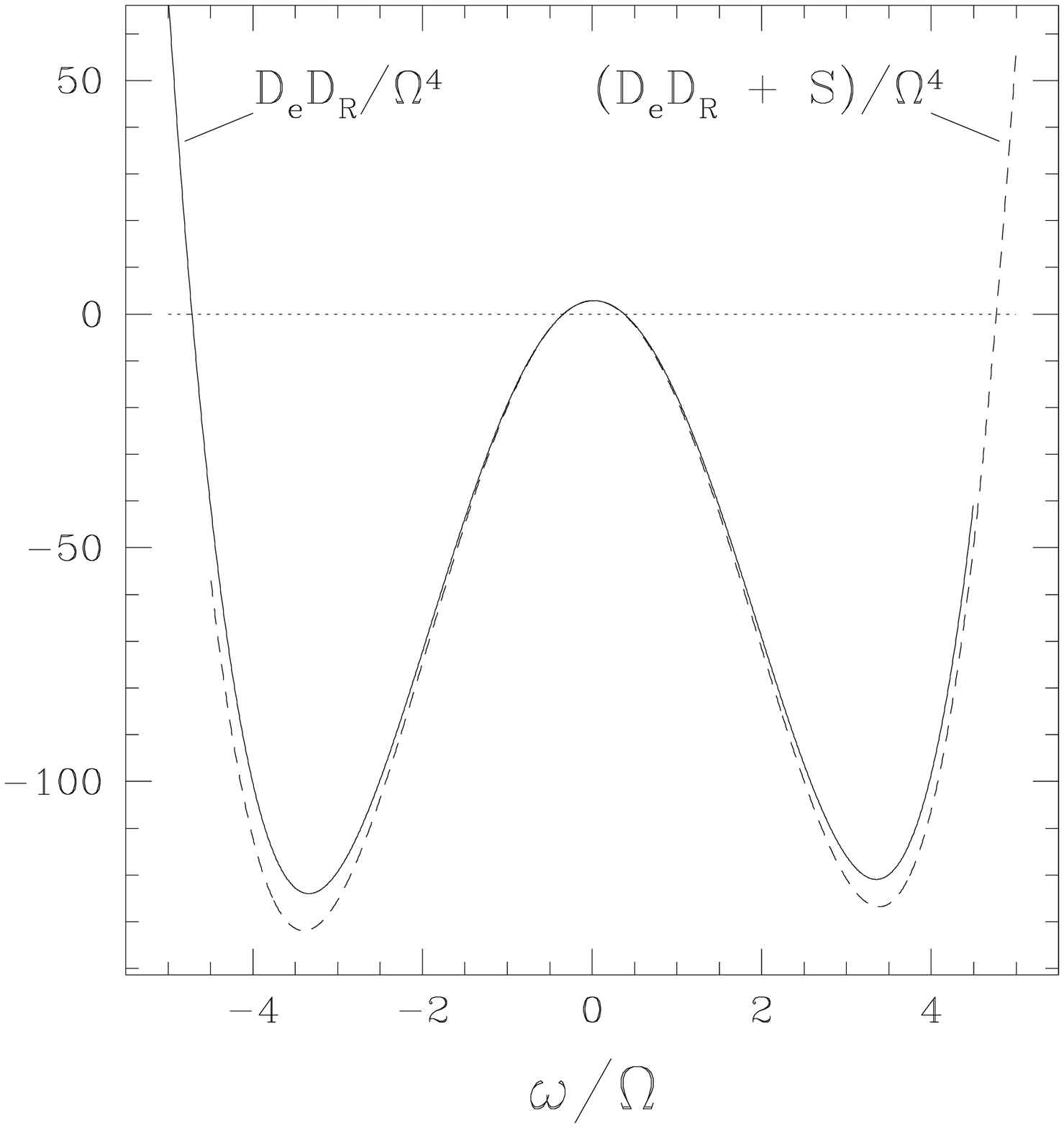,width=3.5in,clip=}
}
\vspace{2mm}
\caption{Plots of $D_{\rm e}D_{\rm R}$ (solid line) and $D_{\rm e}D_{\rm R} + S$
(dashed line),
with $S$ given by equation~(\ref{eq:rl}), $a/v_{\rm A} = 0.2$, 
$v_{\rm A}/v_{\rm rot}= 0.2$, $m=10$, $n=30$, $\ln (r_2/r_1) = 5$,
and $q=5$.
\label{fig:crossings} }
\end{figure*}

When $a/v_{\rm A}$ 
and $v_{\rm A}/v_{\rm rot}$ are small (just
how small will be seen below in the numerical
solutions), both the
approximate roots given in equations~(\ref{eq:ae}) and (\ref{eq:gs})
and the exact roots of equations~(\ref{eq:ldisp1}) and
(\ref{eq:gdisp1}) are real for all $m$ and $n$ if and only if
\begin{equation}
|g| < 2|\Omega| a.
\label{eq:stabcrit} 
\end{equation}
[To be precise, the exact stability
criterion is slightly different from equation~(\ref{eq:stabcrit}) 
due to small corrections associated with nonzero $a/v_{\rm A}$ and
$v_{\rm A}/v_{\rm rot}$.] It can be seen directly from
equation~(\ref{eq:gs}) that the approximate roots are
real for all $m$ and $n$ if and only if equation~(\ref{eq:stabcrit}) 
is satisfied; that the exact roots are also real can be seen
from the following argument. Let
\begin{equation}
D\equiv D_{\rm e}D_{\rm R}+S.
\end{equation}
The four roots of the exact dispersion relation $D=0$ are real when
the plot of $D(\omega)$ passes through zero four times, as in
figure~\ref{fig:crossings}.  When $a \ll v_{\rm A}$ and $v_{\rm
A}\ll v_{\rm rot} $, $D$ always has two local minima at $\omega
\simeq \pm \sqrt{G/2}$ at which $D\simeq -G^2/4$.  Also, $D\rightarrow +\infty$ 
as $|\omega|\rightarrow +\infty$. The question of
stability turns upon whether $D$ is non-negative for some value(s) of
$\omega $ in the interval $ (-\sqrt{G/2}, \sqrt{G/2})$.  When
$a\ll v_{\rm A} $ and $v_{\rm A}\ll v_{\rm rot} $ and
equation~(\ref{eq:stabcrit}) is satisfied, the local maximum of
$D_{\rm e}D_{\rm R}$ at $\omega = \omega_{\rm max} \simeq 2mg\Omega/rG
\in (-\sqrt{G/2}, \sqrt{G/2})$ is positive.  Moreover, 
$S(\omega_{\rm max})$ is negligible compared to the value of
$D_{\rm e}D_{\rm R}$ at $\omega_{\rm max}$ (unless the plasma is
extremely close to marginal stability). Hence, the value of $D$ is
also positive at $\omega_{\rm max}\in (-\sqrt{G/2}, \sqrt{G/2})$, and
the four exact roots are real. On the other hand, when
equation~(\ref{eq:stabcrit}) is not satisfied, the local maximum of
$D_{\rm e}D_{\rm R}$ at $\omega = \omega_{\rm max} $ is negative,
$D_{\rm e}D_{\rm R}$ is negative for all $\omega$ in the
interval $(-\sqrt{G/2},
\sqrt{G/2})$, $|S| \ll |D_{\rm e}D_{\rm R}|$ for all $\omega$ in the
interval $ (-\sqrt{G/2}, \sqrt{G/2})$, $D_{\rm e}D_{\rm R} + S$ is 
negative for all $\omega$ in the interval $(-\sqrt{G/2}, \sqrt{G/2})$, and 
two of the exact roots have nonzero imaginary parts.

The $|m|=1$ roots of equation~(\ref{eq:gdisp1}) are a special case,
since equation~(\ref{eq:gs}) gives a double root at $\omega =
g/2\Omega r$.  In this case, the local maximum of $D_{\rm e}D_{\rm R}$
occurs at $\omega=g/2\Omega r$, and $D_{\rm e}D_{\rm R}=0$ at the
local maximum. It is thus not clear whether the local maximum of
$D_{\rm e}D_{\rm R} + S$ has a positive or negative value. In
appendix~\ref{ap:m1}, a new approximate factorization of
equation~(\ref{eq:gdisp1}) is given for $|m|=1$ and it is shown that the
$|m|=1$ roots are always stable when $a \ll v_{\rm A}$ and $v_{\rm A} \ll v_{\rm
rot}$, whether or not equation~(\ref{eq:stabcrit}) is satisfied.

\begin{figure*}[h]
\centerline{
\psfig{figure=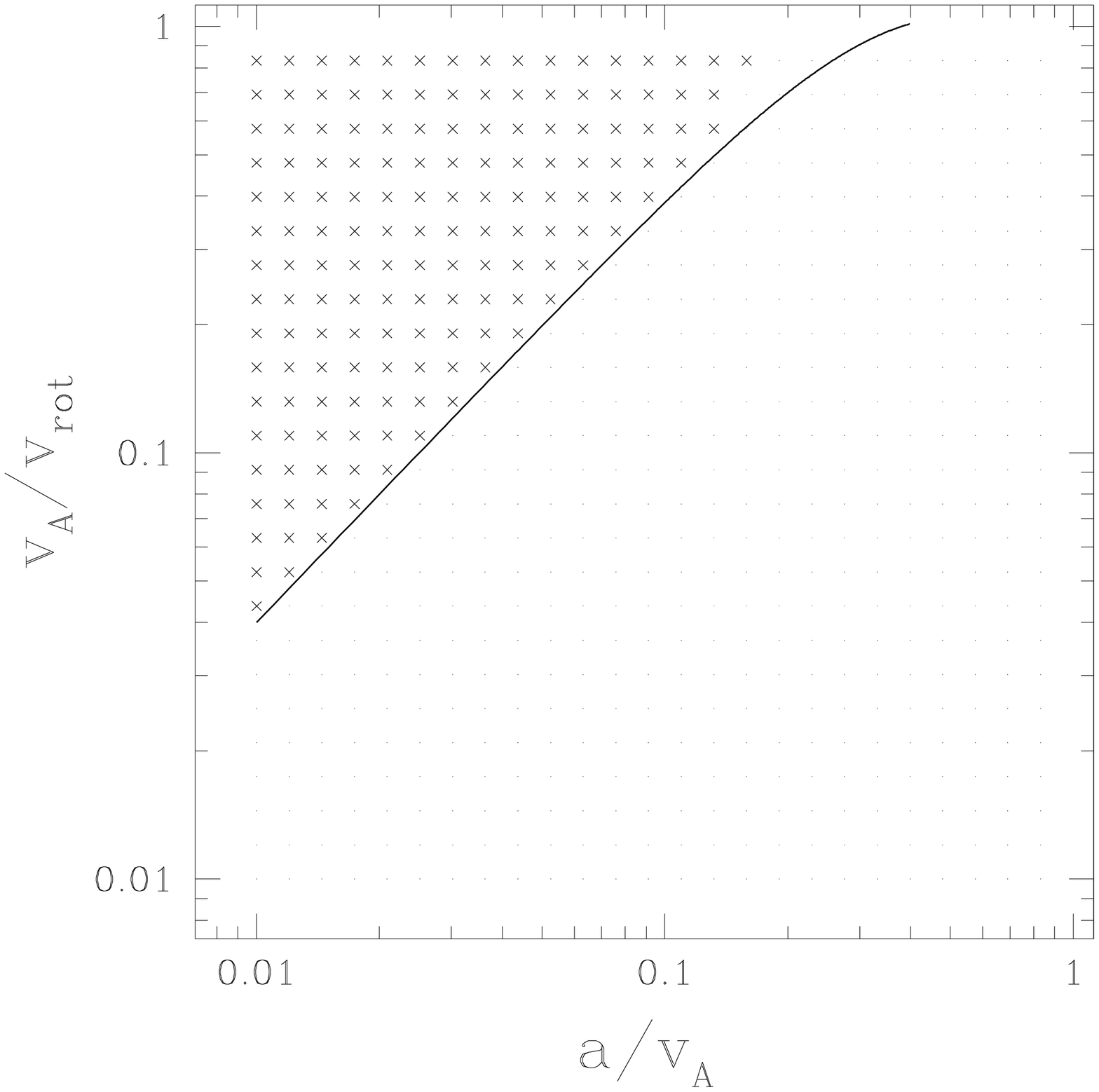,width=3.5in,clip=}
}
\vspace{2mm}
\caption{The crosses indicate combinations of parameters (evaluated in the
dense plasma) for which
the numerical solutions of equations~(\ref{eq:ldisp1}) and (\ref{eq:gdisp1}) 
indicate instability. The dots indicate parameters for which the
numerical solutions are real. The solid line corresponds to
the approximate analytic stability criterion
equation~(\ref{eq:stabcrit2}).
\label{fig:stabcrit} }
\end{figure*}

Figure~\ref{fig:stabcrit} summarizes a stability analysis based upon
numerical solutions of equations~(\ref{eq:ldisp1}) and
(\ref{eq:gdisp1}) for values of $a/v_{\rm A}$ and $v_{\rm A}/v_{\rm
rot}$ ranging from 0.01 to 1. A combination of parameters $(a/v_{\rm
A})$ and $(v_{\rm A}/v_{\rm rot})$ is said to be stable if the
imaginary part of each of the numerically determined roots is less
than $10^{-10}$ times the real part of the same root for all values of
$m$ from 1 to 100, and in the case of the roots to
equation~(\ref{eq:ldisp1}), for all values of $n$ from 1 to 500, where
$\ln(r_2/r_1)$ is taken to be 5, and $q$ is taken to be 5.  The roots
are determined using an eigenvalue method (Press et~al 1992) at
double precision. Equation~(\ref{eq:stabcrit}) describes
the stability of the exact roots quite accurately provided $v_{\rm
A}/v_{\rm rot} \lesssim 0.5$.  For larger values of $v_{\rm A}/v_{\rm
rot}$, equation~(\ref{eq:stabcrit}) underestimates stability.

For any given values of $a$, $\Omega$, and $g$, the most unstable mode
has zero Lagrangian total-pressure perturbation everywhere and 
$\zeta = m^2 -1$. Upon setting
$d \omega_I/dm = 0$, where $\omega_I$ is the
imaginary part of $\omega$, one can show using
equation~(\ref{eq:gs}) that the
maximum growth rate $\gamma_{\rm max}$ occurs at
\begin{equation}
m_{\rm max} \simeq \pm \frac{v_{\rm rot}}{v_{\rm A}}
\sqrt{ -6 + \sqrt{4 + \frac{8g^2}{\Omega^2 a^2}}}
\label{eq:mmax} 
\end{equation} 
and is given by
\begin{equation}
\gamma_{\rm max} \simeq \frac{g}{v_{\rm A}} f(\Omega^2 a^2/g^2),
\label{eq:gammamax} 
\end{equation} 
where $f$ is a function that increases monotonically from 0 to 1 as
its argument $\Omega^2 a^2/g^2$ is decreased from $0.25$ (marginal
stability) towards~0.  Because of equation~(\ref{eq:gva}),
$\gamma_{\rm max} \sim v_{\rm A}/r$ when the plasma is not very near
marginal stability. When $|g|> 2|\Omega| a$, modes with $\zeta = m^2 -1$
as well as modes with $\zeta = p_n +m^2 -1$ 
are unstable only
when $\zeta < \zeta_{\rm crit}$, where
\begin{equation} 
\zeta_{\rm crit} = \frac{r^2 ( g^2 - 4\Omega^2 a^2)}{a^2 v_{\rm A}^2}
\label{eq:zetacrit} .
\end{equation} 

\section{Discussion of stability criterion and
eigenmodes}
\label{sec:stability} 

In the absence of background flow, the stability of an equilibrium to
a small-amplitude displacement ${\bf \xi}({\bf x})$ is determined by
whether ${\bf \xi}({\bf x})$ leads to an increase or decrease in
$\delta W$, where $\delta W$ is the sum of the perturbed magnetic,
internal, and gravitational potential energies of the plasma
(Bernstein et~al 1958). If $\delta W$ is positive, then the mode is
stable---if the displacement increases with time
initially, the initial bulk-motion kinetic energy present in the mode
is converted into these other types of energy, and the plasma proceeds
to oscillate stably. If $\delta W$ is negative, then the magnetic,
internal, and gravitational energies of the plasma can be tapped to
drive an instability.  In an equilibrium with flow, the kinetic energy of the
flow is altered by plasma displacements, and stability is no longer
determined solely by the increments to the magnetic, internal, and
gravitational potential energies. The perturbed kinetic energy depends
upon both the displacement ${\bf \xi}({\bf x})$ and its time
derivative $\partial{\bf \xi}/\partial t$, which gives rise to the
phenomenon of overstability, in which both the real and imaginary
parts of $\omega$ are nonzero, and which makes the stability analysis
more complicated than in the static case (Frieman \& Rotenberg 1960).

In this section, the physical basis of the stability criterion is
explained using the properties of the eigenmodes. For any set of
plasma parameters, the most unstable $k_z=0$ modes are those described
by equation~(\ref{eq:gdisp1}) [as opposed to those described by
equation~(\ref{eq:ldisp1})], for which the Lagrangian total-pressure
perturbation vanishes everywhere. It is these modes that will be
discussed in this section. The dispersion relation given by
equation~(\ref{eq:gdisp1}) has two branches, described approximately
by equations~(\ref{eq:ae}) and (\ref{eq:gs}).  Equation~(\ref{eq:ae})
is an epicyclic-Alfv\'enic branch.  When $m \sim {\cal O}(1)$, one
finds that $\omega = \pm 2\Omega$, $u_{1\theta} \simeq \mp i u_{1r}$,
and $b\simeq -(\pm m+1)a^2/v_{\rm A}^2$.  These circularly polarized
epicyclic modes, which involve primarily a balance between inertia and
the Coriolis force, are stable for all plasma parameters: when a fluid
element in the dense plasma of figure~\ref{fig:equil_picture} tries to
sink towards $r=0$, the Coriolis force pushes it in the
$+\hat{\theta}$ direction (for positive $\Omega$); as it moves in the
$+\hat{\theta}$ direction, the Coriolis force pushes it back towards
larger~$r$. For $m^2 \gg v_{\rm rot}^2/v_{\rm A}^2$, the modes are
stable Alfv\'en waves, with $\omega = \pm mv_{\rm A}/r$,
$|u_{1\theta}| \ll |u_{1r}|$, and $b\simeq -a^2/v_{\rm A}^2$. The
dominant force across field lines for these modes is magnetic tension.

Equation~(\ref{eq:gs}) is the Rayleigh-Taylor/sound-wave branch of the
dispersion relation. It is shown in appendix~\ref{ap:eigenmodes} that
the eigenmodes corresponding to equation~(\ref{eq:gs}) satisfy $|u_{1\theta}| \gg
|u_{1r}|$, which indicates that these modes involve comparatively
little field-line bending. For $m \sim {\cal O}(1)$, $\omega \ll \Omega$. In
the frame rotating with the plasma, the momentum equation at first
order in perturbation amplitude can be written $-i\omega \rho_0{\bf
u}_1 = -2\rho_0 \Omega {\hat z} \times {\bf u}_1 + \dots$, where only
the Coriolis force ${\bf F}_{\rm C}$ has been written explicitly on
the right-hand side.  Since $\omega\ll \Omega$ and ${\bf u}_1$ is not
along $\hat{z}$, some force must approximately cancel the Coriolis
force.  For $m\sim {\cal O}(1)$, the near cancellation is provided by
the gravitational force ${\bf F}_g$.  Upon defining $F_{g,\parallel} =
{\bf F}_g\cdot {\bf B}/B_0$ and $F_{\rm C,\parallel} = {\bf F}_{\rm C}
\cdot {\bf B}/B_0$, and using equation~(\ref{eq:b1r}) to eliminate
$B_{1r}$, one obtains
\begin{eqnarray} 
F_{g,\parallel} & = & \frac{\rho_0 mg u_{1r}}{\omega r}, \mbox{ \hspace{0.3cm} and}
\label{eq:fgp} \\
F_{\rm C,\parallel} & = & - 2 \rho_0 \Omega u_{1r}.
\label{eq:fcp} 
\end{eqnarray}
It is shown in appendix~\ref{ap:eigenmodes}  that $\rho_1 \simeq
\rho_0 m u_{1\theta} /\omega r$ (that is, the density perturbation
is dominated by the compression associated with $u_{1\theta}$), so that 
\begin{equation}
\hat{r} \cdot {\bf F}_g \simeq -\frac{\rho_0 m g u_{1\theta}}{\omega r},
\label{eq:fgr} 
\end{equation} 
while
\begin{equation}
\hat{r} \cdot {\bf F}_{\rm C} = 2\rho_0\Omega u_{1\theta}.
\label{eq:fcr} 
\end{equation} 
The phase velocity of a low-$m$ mode in the $\hat{\theta}$ direction,
$v_{\rm phase} = \omega r /m$, is set by the requirement that ${\bf
F}_{\rm C}$ and ${\bf F}_g$ approximately cancel, which yields $v_{\rm
phase} \simeq g/2\Omega$, or $\omega \simeq mg/2\Omega r$. This phase velocity
is in the eastwards direction when $g>0$---i.e., in the direction of rotation
($\omega/ m \Omega > 0$).  Because of the additional forces in the
plasma, which are much smaller than either ${\bf F}_g$ or ${\bf F}_{\rm C}$,
the exact phase velocity will be either a little larger than
$|g/2\Omega|$, in which case $|F_g| < |F_{\rm C}|$, or a little
smaller than $|g/2\Omega|$, in which case $|F_g| > |F_{\rm C}|$.

It will prove useful to consider the radial momentum equation in
somewhat greater detail.  It is shown in appendix~\ref{ap:eigenmodes}
that $|b| \ll 1$ for the eigenmodes of equation~(\ref{eq:gs}) with
$m\sim {\cal O}(1)$.  This is related to the fact that the Lagrangian
perturbation of the magnetic pressure, which is $\propto du_{1r}/dr$,
is cancelled by the Lagrangian perturbation of the much smaller
thermal pressure. (Although $|u_{1\theta}/u_{1r}|$ is large, it is not
large enough to enable $|b|$ to be $\gtrsim 1$.) With the use of the
relation $|b| \ll 1$, the vanishing of the Lagrangian total-pressure
perturbation, equations~(\ref{eq:b1r}) and (\ref{eq:b1t}), and the
equation $\rho_1 = - \rho_0 \nabla \cdot {\bf \xi} - {\bf \xi} \cdot
\nabla \rho_0$ (Frieman \& Rotenberg 1960), the radial component of
the momentum equation can be written
\begin{equation}
- i \omega \rho_0 u_{1r} \simeq \frac{\rho_0 u_{1\theta}}{\omega}
\left(2\Omega \omega - \frac{mg}{r}\right) 
- \frac{i u_{1r} \rho_0 (m^2 -1) v_{\rm A}^2}{\omega r^2}.
\label{eq:arm} 
\end{equation} 
The $(\rho_0 u_{1\theta}/\omega) ( 2\Omega \omega - mg/r)$ term is the
radial component of what will be called the net gravitational force,
${\bf F}_g + {\bf F}_{\rm C}$.  The $-i u_{1r} \rho_0 (m^2 -1 ) v_{\rm
A}^2/\omega r^2$ term is the effective magnetic tension, which acts to
restore a radially displaced fluid element in the dense plasma to its
initial position.  In order for a mode with $\omega \simeq mg/2\Omega
r$ to propagate, the radial components of the effective tension and
the net gravitational force must approximately cancel; otherwise,
solving for $\omega$ yields $|\omega| \gtrsim v_{\rm A}/r$, which is
$\gg mg/2\Omega r$ for $m \sim {\cal O}(1)$.

\begin{figure*}[h]
\vspace{10cm}
\includegraphics{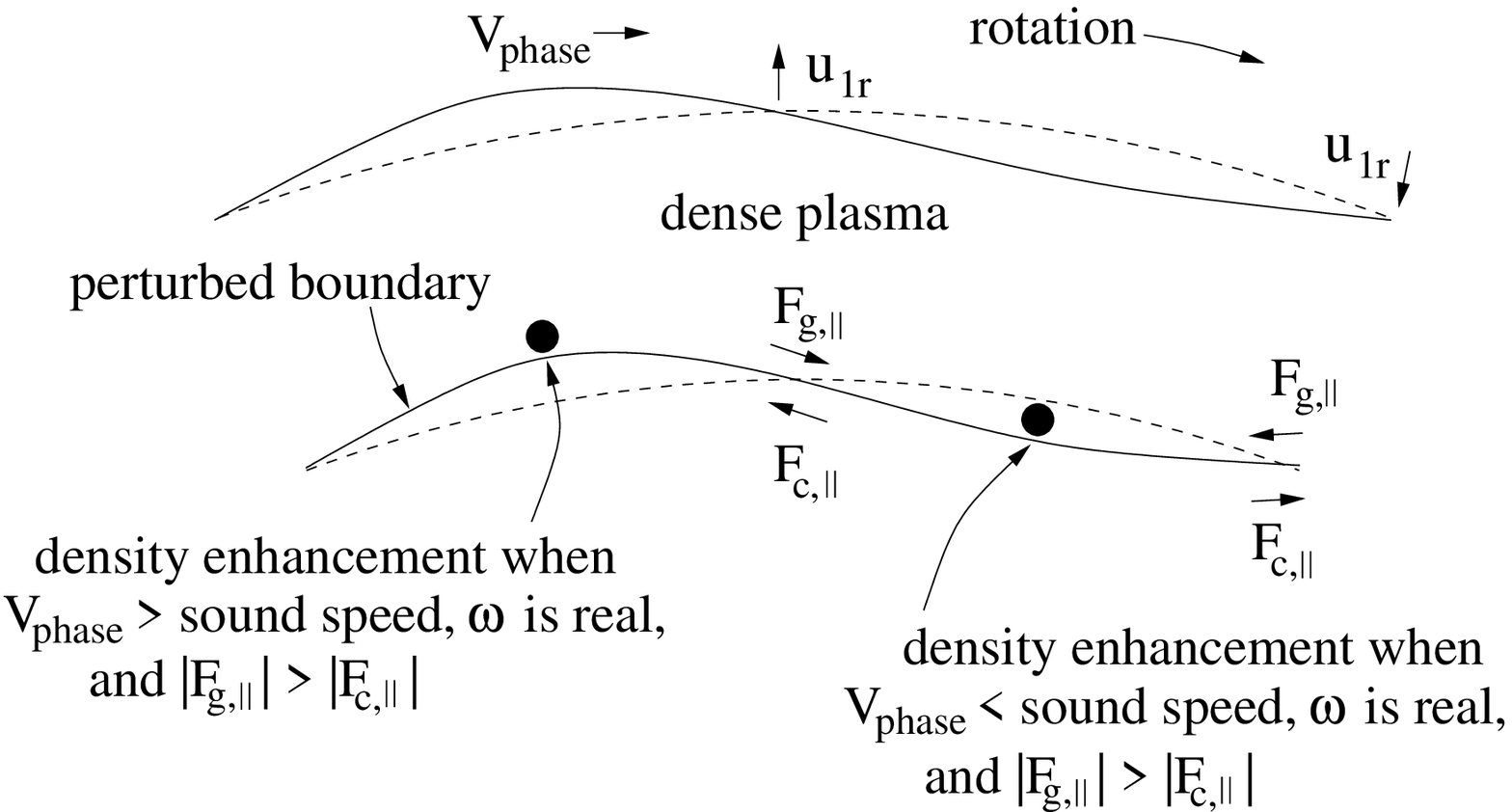}
\caption{Eigenmodes corresponding to equation~(\ref{eq:gs}) with
$m \sim {\cal O}(1)$ and a Lagrangian total-pressure perturbation that
vanishes everywhere. $F_{g,\parallel}$ and $F_{\rm C,\parallel}$ are the
components of the gravitational and Coriolis forces along the
perturbed magnetic field lines in the dense plasma. When $\omega$ is
real and $v_{\rm phase} < a$, the plasma is compressed at the points
towards which $F_{g,\parallel} + F_{\rm C,\parallel}$ converges, so
that the perturbed pressure force opposes $F_{g,\parallel} + F_{\rm
C,\parallel}$.  When $\omega$ is real and $v_{\rm phase} > a$, the
plasma responds inertially and is compressed at the points away from
which $F_{g,\parallel} + F_{\rm C,\parallel}$ 
diverges. In the latter case, magnetic tension
is unable to balance the perturbed radial gravitational force, and no
stable mode exists.
\label{fig:eigen1}  }
\end{figure*}

Because gravity and the Coriolis force nearly cancel, the low-$m$
eigenmodes of equation~(\ref{eq:gs}) do not enjoy the full stabilizing
effect that the Coriolis force has on epicyclic modes, nor the full
destabilizing effect that gravity has on Rayleigh-Taylor modes in a
static plasma. The physics behind the stability criterion can be
understood by considering a mode with $m \sim {\cal O}(1)$ and
$|v_{\rm phase}|$ slightly less than $|g/2\Omega|$, which implies that
$|F_g|$ is slightly bigger than $|F_{\rm C}|$ and that the net
gravitational force acts in the same direction as gravity. If $|v_{\rm
phase}| < a$, then the dense plasma reacts in a way that is similar to
how it would react if $v_{\rm phase}$ were~0: the maximum density
enhancements occur at those points towards which the parallel
component of the net gravitational force converges---the troughs of
the perturbed magnetic field lines, as in figure~\ref{fig:eigen1}.
The induced pressure force opposes the parallel component of the net
gravitational force.  In the radial direction, the net gravitational
force and the effective magnetic tension oppose each other and can
therefore approximately cancel, as required for a mode with $\omega
\simeq mg/2\Omega r$.  On the other hand, if $|v_{\rm phase}| > a$,
then sound waves can not propagate information about the net gravity
upstream of the wave. At any instant in time, fluid elements at the
troughs of the perturbed field lines can not tell that the net
gravitational force is trying to compress them, and the plasma reacts
inertially to the parallel component of the net gravitational force.
Plasma is accelerated towards positive $\hat{\theta}$ at those points
where $B_{1r}/B_0 < 0$, and the resulting azimuthal compression causes
the density enhancements to occur at the crests of the field lines
when $\omega$ is real, as in figure~\ref{fig:eigen1}. [This point can
also be seen with the aid of equation~(\ref{eq:aparallel_mom}) in
appendix~\ref{ap:eigenmodes}.]  In this case, the net gravitational
force in the radial direction acts in the same direction as the
effective magnetic tension, providing much more radial force than is
consistent with $\omega \simeq mg/2\Omega r$, and no stable mode at
approximately that frequency exists.

Overstable modes, however, do exist.  One way of thinking
about this is the following.  Suppose a low-$m$ mode were to start
propagating with a phase velocity greater than $a$ and peak density
enhancements at the crests of waves (as would be the case for real
$\omega$).  The radial component of the net gravitational force and
magnetic tension would then provide much more force than required to
return the displaced plasma to its initial position.  A fluid element
displaced a distance $\xi_{r0}$ towards $+\hat{r}$ would in response
pass through its initial position and then move a distance $>
\xi_{r0}$ towards $-\hat{r}$, leading to an oscillation of growing
amplitude.  If $\omega_I \neq 0$, where $\omega_I$ and $\omega _R$ are
the imaginary and real parts of $\omega$, respectively, then the phase
relation between $\rho_1$ and $u_{1r}$ is not as depicted in
figure~\ref{fig:eigen1}.  When $m \sim{\cal O}(1)$, $|\omega_R| \gg
|\omega_I|$ as in equation~(\ref{eq:gs}).  $F_{g,\parallel}$ and
$F_{\rm C,\parallel}$ are then effectively sine waves of equal
amplitude that are nearly $180^\circ$ out of phase, 
and their sum is about 90~degrees out of
phase with each of them. The locations at which $\rho_1$ obtains its
peak magnitude are then approximately the steepest points on the
perturbed magnetic field lines, in phase with $B_{1r}$ and
$u_{1r}$. The peak magnitude of the radial component of the net
gravitational force is then about 90 degrees out of phase with
$\rho_1$ and coincides with the troughs and crests of field lines. The
radial component of the net gravity is then approximately cancelled by
the effective magnetic tension.

When $|g| > 2|\Omega|a$, modes are stabilized when $\zeta$ is
increased above $\zeta_{\rm crit}$ because pressure begins to dominate
over gravity in the dynamics along the magnetic field lines, changing
the modes from Rayleigh-Taylor instabilities into something akin to 
sound waves.  It is shown in appendix~\ref{ap:eigenmodes} that
$|F_{g,\parallel}| > |F_{p, \parallel}|$ when $\zeta < \zeta_{\rm
crit}$ and $|g| > 2|\Omega| a$, where $F_{p,\parallel}$ is the
component of the pressure force along the perturbed magnetic field
lines in the dense plasma.  It is also shown that as $\zeta$ is
increased towards $\zeta_{\rm crit}$,
$|F_{p,\parallel}|/|F_{g,\parallel}|\rightarrow 1$. For
$\zeta>\zeta_{\rm crit}$, $|F_{p,\parallel}| > |F_{g, \parallel}|$ for
all $m$ with one exception: $\zeta_{\rm crit} < \zeta < g^2 r^2/a^2
v_{\rm A}^2$ on the branch of the dispersion relation with $\pm \Omega
<0$. This exception corresponds to the segment of the plot of
$\omega_R$ in the upper-left panel of figure~\ref{fig:comparison} just
to the right of $\zeta_{\rm crit}$ where the curve bifurcates, between
the bifurcation point and where the plot of $\omega_R$ intersects
$\omega_R = 0$; the plot of $\omega_R$
bifurcates at $m=49$, and the exception corresponds to the smaller
value of $\omega$ at $m=50$.

To obtain further insight into the stability criterion, it is helpful
to consider some limiting cases.  Of course, if there is neither
rotation nor magnetic field, the equilibrium of
figure~\ref{fig:equil_picture} is unstable to Rayleigh-Taylor modes as
in the classic calculation of a dense fluid suspended above
low-density fluid in slab geometry (Rayleigh 1883).  If the orthogonal
magnetic fields of figure~\ref{fig:equil_picture} are present but the
plasma isn't rotating, the situation is similar to the stationary
plane-parallel problem solved by Gratton, Gratton, \& Gonzalez (1988)
and Gonzalez \& Gratton (1990). These authors discussed a number of
cases, including that of a plasma at $y>0$ suspended above vacuum at
$y<0$ with gravity pointing in the $-\hat{y}$ direction, all
equilibrium quantities being functions of $y$ alone, the magnetic
field in the plasma pointing in the $x$ direction, and the magnetic
field in the vacuum pointing in the $z$ direction. They found that the
orthogonal magnetic fields act to stabilize the plasma, since the
interface at $y=0$ can not be perturbed without bending either the
magnetic field in the plasma or the magnetic field in the vacuum.
Compressibility, however, acts to destabilize
the plasma; if the plasma is strongly compressed along the magnetic
field, the gravitational force acting on regions of enhanced density
can overcome magnetic tension. When the wave vector ${\bf k}$ of the
perturbation in the $x-z$ plane is oriented along the magnetic field in
the plasma at $y>0$, magnetic tension on its own is not sufficient to
stabilize small-scale modes: as $a\rightarrow 0$, the maximum unstable
wavenumber approaches $\infty$. Since thermal pressure limits the 
compressibility of the plasma along the magnetic field, 
magnetic tension and pressure together stabilize modes with $k^2> k_{\rm crit}^2$,
where $k_{\rm crit}^2 = g^2(v_{\rm A}^2 + a^2)/a^2 v_{\rm A}^4$.  It
can be shown that the condition $|k|>k_{\rm crit}$ corresponds to
pressure dominating over gravity in the dynamics along the perturbed
magnetic field lines, in analogy to the stabilization of the modes
described by equation~(\ref{eq:gs}) when $\zeta> \zeta_{\rm crit}$.
On the other hand, sufficiently large-wavelength modes ($|k|< k_{\rm
crit}$) are always unstable in the magnetized slab, whereas all
$k_z=0$ modes are stable in the rotating magnetized cylinder when $|g|
< 2|\Omega|a$. 

If there is rotation but no magnetic field, the plasma has a set
of normal modes analogous to those described by
equation~(\ref{eq:gdisp1}) (i.e., $k_z=0$ and zero Lagrangian pressure
perturbation everywhere), with frequencies
\begin{equation}
\omega = \Omega \pm \sqrt{\Omega^2 - \frac{mg}{r}}
\label{eq:oh1} 
\end{equation}
and
\begin{equation}
\omega = -\Omega \pm \sqrt{\Omega^2 + \frac{mg}{r}}.
\label{eq:oh2} 
\end{equation} 
These modes are circularly polarized ($u_{1\theta} = \pm u_{1r}$),
stable at long wavelengths ($|m| < \Omega^2r/g$), unstable when $m >
\Omega^2 r/g$ in equation~(\ref{eq:oh1}), and unstable when $m < -
\Omega^2 r/g$ in equation~(\ref{eq:oh2}). As $|m| \rightarrow \infty$, the growth time of
unstable modes ($\sim 1/\sqrt{mg/r}$) is much shorter than the
rotation period, and rotation can be neglected. The introduction of an
azimuthal magnetic field into the dense plasma has a stabilizing
influence insofar as magnetic tension eliminates these large-$|m|$
instabilities with $|u_{1\theta}| = |u_{1r}|$.

The magnetic field in the equilibrium of
figure~\ref{fig:equil_picture}, however,
can also be destabilizing: as the field
strength is increased, the plasma becomes increasingly supported by
magnetic pressure, thereby increasing the effective gravity~$g$.
With the use of equation~(\ref{eq:gva}), the stability criterion
can be rewritten as an upper limit on the Alfv\'en speed:
\begin{equation}
v_{\rm A}^2 < \left(\frac{2}{\alpha}\right) a v_{\rm rot}
\label{eq:stabcrit2} 
\end{equation} 
(where $a$, $v_{\rm A}$, and $v_{\rm rot}$ are evaluated in the dense plasma).
In the low-$\beta$ limit, pressure balance requires that the
magnetic-field strength $B_{\rm vert}$ in the low-density plasma at
$r< r_1$ be approximately equal to the field strength in the dense
plasma near $r=r_1$, $\overline{ B}$.  Assuming a dense plasma of pure
hydrogen with $q=5$, $\gamma=5/3$, $\beta\ll 1$, $v_{\rm rot} = 150$
km/s [from the Galactic rotation curve at $r= 150$ pc (Ruzmaikin \&
Shukurov 1981)], and $T= 70$ K and $n = 10^4 \mbox{ cm}^{-3}$ near
$r=r_1$, one can rewrite equation~(\ref{eq:stabcrit2}) as
\begin{equation}
B_{\rm vert} < 1.1 \mbox{ mG}.
\label{eq:stabcrit3} 
\end{equation}
Although it is tempting to infer from equation~(\ref{eq:stabcrit3}) that
mG vertical magnetic fields at the Galactic center
can be stably confined by a ring of
horizontal magnetic fields threading molecular material,
further study is needed to extend the present analysis to
more realistic equilibria.

\section{Other types of instabilities}
\label{sec:local} 

Sections \ref{sec:diffeq} through \ref{sec:stability} treated the
global stability of the equilibrium pictured in
figure~\ref{fig:equil_picture} to $k_z=0$ modes. This section
addresses three other classes of instabilities: (1)~local interchange
and non-axisymmetric quasi-interchange modes in a uniformly rotating
plasma with azimuthal magnetic field, (2)~global instabilities in the
1D equilibrium of figure~\ref{fig:equil_picture} 
with $k_z \neq 0$, and (3)~quasi-interchange and
magneto-rotational instabilities in a two-dimensional differentially
rotating equilibrium such as the one pictured in
figure~\ref{fig:equil}.

Local modes in a plasma with an azimuthal magnetic field that rotates
uniformly about the $z$ axis have been treated by a number of authors.
For the constant-$\beta$ equilibrium profile of section~\ref{sec:eq},
local axisymmetric interchange instabilities are
suppressed primarily because of the strongly stabilizing
specific-angular-momentum profile when $a\ll v_{\rm A} \ll v_{\rm
rot}$ (Newcomb 1962, Acheson \& Gibbons 1978, Rogers \& Sonnerup 1986,
Ferriere et~al~1999).  Such modes are further stabilized since the
specific entropy and magnetic flux per unit mass of an azimuthal flux
loop $B/\rho r$ are taken to increase with~$r$ in the constant-$\beta$
equilibrium of section~\ref{sec:bc}.  The results of Ferriere et~al
(1999) on a wide class of one-dimensional equilibria can be used to
show that local non-axisymmetric quasi-interchange modes 
in a uniformly rotating plasma
with azimuthal magnetic field are stable when two conditions are met
assuming $a\ll v_{\rm A} \ll v_{\rm rot}$: (1) the ``magnetic
Brunt-V\"{a}is\"{a}l\"{a}'' frequency $\omega_0$ must be real, and (2)
$|g| < 2|\Omega| a$ (to lowest order in $a/v_{\rm A}$ and $v_{\rm
A}/v_{\rm rot})$.  For the constant-$\beta$ equilibrium of
section~\ref{sec:bc}, $\omega_0^2 \simeq qg/2r $ and the first
condition is met.  The second condition is the same as
equation~(\ref{eq:stabcrit}), and thus when the $k_z=0$ modes are
stable, so are the quasi-interchange modes.  The quasi-interchange
modes have $\omega \simeq mg/2\Omega r$, and are stable (provided
$\omega_0$ is real) when their phase velocity along the magnetic field
is less than the sound speed, just as for $k_z=0$ modes with $m\sim
{\cal O}(1)$.

\begin{figure*}[h]
\vspace{10cm}
\includegraphics{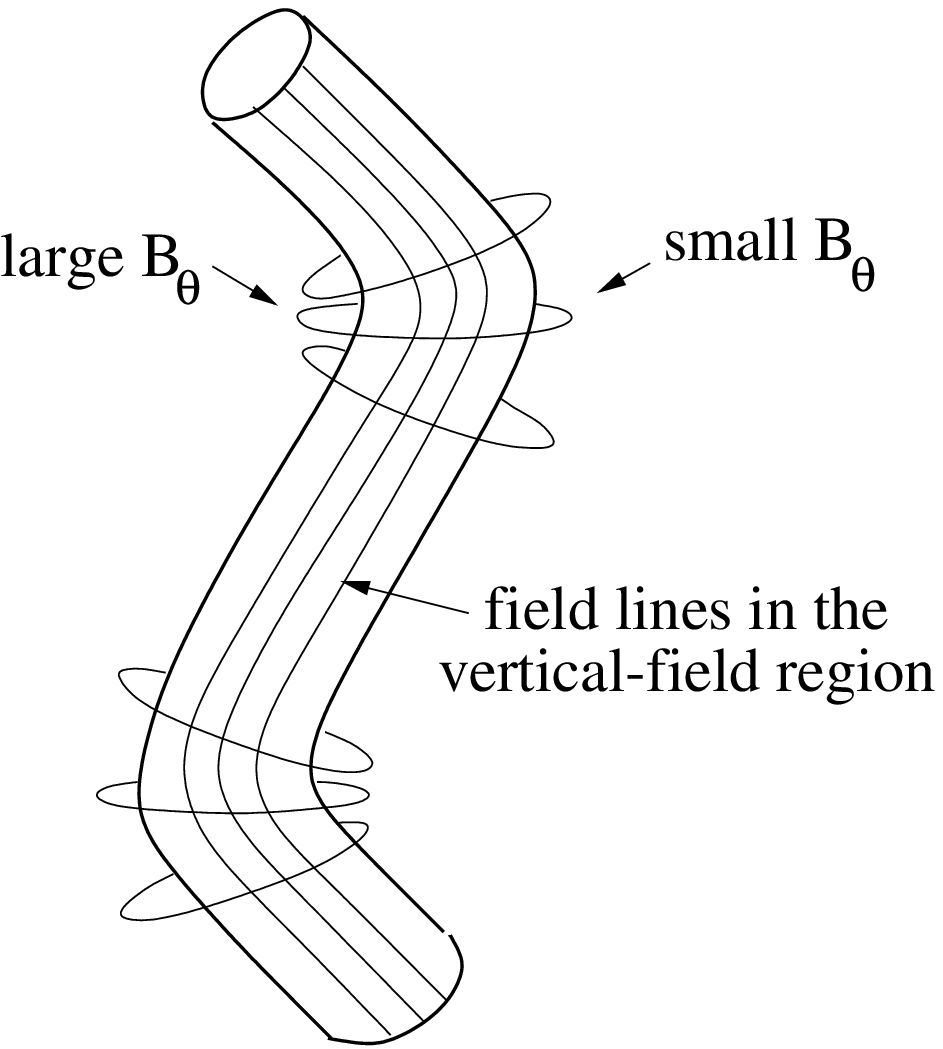}
\caption{An $m=1$ external kink 
mode. \label{fig:kink}   }
\end{figure*}

Global modes in the equilibrium of figure~\ref{fig:equil_picture} 
with $k_z \neq 0$ and $m\sim {\cal O} (1)$ are analogous
to the external kink instability in static plasmas. An $m=1$ mode is
illustrated in figure~\ref{fig:kink} [figure 9.23 of Freidberg
(1987)]. Because the strength of the azimuthal field increases where
the field lines are compressed, the Lorentz force of the azimuthal
field acts to increase the perturbation.  On the other hand, the
magnetic tension of the axial magnetic field opposes the
perturbation. In the static reversed field pinch, in which $B_\theta
\sim B_z$, the azimuthal magnetic field wins out over the axial field
and the mode is unstable when the axial wavelength $k_z^{-1}$ is
greater than roughly the radius of the plasma cross section (Friedberg
1987). How this instability mechanism affects the stability criterion
of the rotating equilibrium of figure~\ref{fig:equil_picture} is left
for future work.

Newcomb (1961) showed that the convective stability criterion for a
compressible stratified magnetized slab when $k_\parallel =0$ is less
stringent than the stability criterion in the limit $k_\parallel
\rightarrow 0$, where $k_\parallel$ is the component of the wave
vector along the magnetic field.  This is because when $k_\parallel =
0$, Newcomb's type~2 modes, which involve motion primarily along the
magnetic field, become pure translations which are neutrally stable
($\omega = 0$). The $k_\parallel = 0$ stability criterion is thus the
one that applies to Newcomb's type~1 modes, which involve motion
primarily across field lines. On the other hand, when $k_\parallel $
is small but nonzero, the type~2 modes allow destabilizing
compressions of the plasma along the magnetic field and can be
unstable even when type~1 modes are stable.  The distinction between
$k_\parallel = 0$ and $k_\parallel \rightarrow 0$ in Newcomb's work is
not directly analogous to the distinction between $k_z =0 $ and
$k_z\rightarrow 0$ in the equilibrium of
figure~\ref{fig:equil_picture}, in that $k_\parallel$ corresponds to
$m$ since it is the dense plasma that dominates the dynamics of the
unstable modes.

The two-dimensional equilibrium of figure~\ref{fig:equil} is subject
to instabilities not present in the one-dimensional infinite-cylinder
equilibrium. If the vertical (along $z$) stratification does not
satisfy the Schwarzchild criterion for a static non-magnetized gas,
then the plasma is unstable to local non-axisymmetric
quasi-interchange modes with $k_r \gg k_z$ and $k_r r/ m \gg 1$
(Terquem \& Papaloizou 1996). The displacements in such modes are
essentially parallel to the spin axis, and thus the Coriolis force
vanishes, preventing the type of rotational stabilization that occurs
for quasi-interchange modes in the one-dimensional infinite-cylinder
equilibrium.  If there is differential rotation with $d|\Omega| /dr <
0$, the plasma is subject to the magnetorotational instability (MRI)
(Ogilvie \& Pringle 1996, Terquem \& Papaloizou 1996, Balbus \& Hawley
1998).  Global three-dimensional numerical simulations suggest that
turbulence excited by the MRI will be present quite generally in astrophysical
disks (Hawley, Balbus, \& Stone 2001). It is thus interesting to ask
how the stability of the sharp boundary in figure~\ref{fig:equil} is
affected by differential rotation in a quiescent equilibrium flow as
well as by differential rotation with turbulence.  These questions
are left for future work.

\section{Estimates of field strengths in radio filaments}
\label{sec:review} 

Lang et~al.\ (1999b) found that if the magnetic energy density in the
Northern Thread is equal to the relativistic-particle energy density,
and if the relativistic protons and electrons have equal energies,
then the field strength is $\simeq 140 \mu$G. Comparable equipartition
field strengths have been found for the Radio Arc (Tsuboi et~al.\
1986, Yusef-Zadeh \& Morris 1987b, Reich 1994). However, the
equipartition estimate may lead to an underestimate of the field
strength, particularly since the geometry of the filaments allows
relativistic particles to stream directly out of the Galactic-center
region.

All of the radio filaments that have been studied in detail near the
Galactic center appear to be interacting with at least one randomly
moving molecular cloud, and yet most are undeformed at the sites of
interaction (Morris \& Serabyn 1996). If the magnetic field were
dynamically insignificant, one would expect the turbulent motions of
the clouds to cause turbulent bulk motions in the hot plasma that
would deform the frozen-in magnetic field lines.  The lack of
deformation gives a lower limit on the field strength in the filaments
that depends upon how many independently moving clouds (or clumps
within a cloud) interact with a single filament.

\begin{figure*}[h]
\vspace{10cm}
\includegraphics{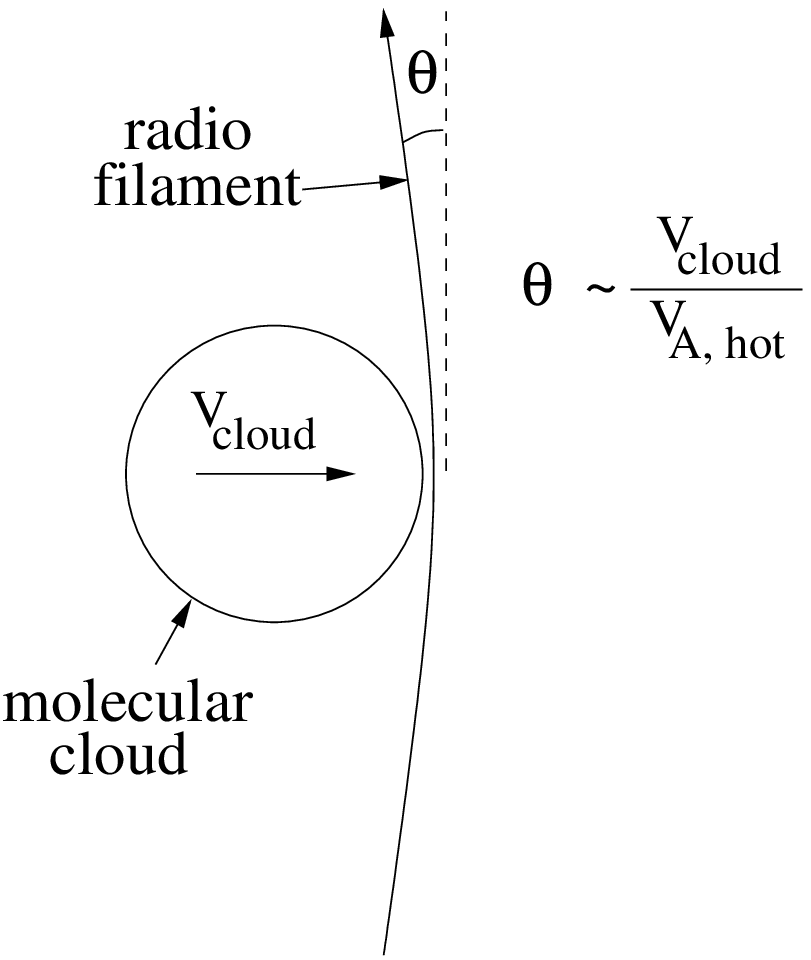}
\caption{Because the field lines in the hot low-density plasma are not anchored
above or below the Galactic plane, they remain almost
straight if they interact with a cloud at only a single point along
their length provided the Alfv\'en speed in the hot plasma $v_{A, {\rm hot}}$
is $\gg v_{\rm cloud}$. \label{fig:deformation}   }
\end{figure*}

If a vertical filament interacts with a single molecular cloud at a
single point along its length, then the lower limit on its field
strength is fairly small.  Since the vertical field $B_{\rm vert}$
threads the very tenuous hot plasma, there is very little mass to
anchor the ends of the vertical field line above and below the
Galactic disk. Any deformation in the field line in the galactic plane
by an interaction with a molecular cloud is carried away along the
field line at the Alfv\'en speed of the hot medium, $v_{A, {\rm
hot}}$.  In order for the visible deformations in the filaments to be
small, the Alfv\'en speed in the hot medium must be much larger than
the turbulent speed of the the clouds $v_{\rm cloud}$ with respect to
the ends of the filament.  The angle at which a field line is bent
will be $\sim v_{\rm cloud}/v_{A, {\rm hot}}$, as depicted in
figure~\ref{fig:deformation}.  The condition $v_{A, {\rm hot}} \gg
v_{\rm cloud}$ can be written
\begin{equation}
B_{\rm vert} \gg   n_{\rm hot}^{1/2} \left(\frac{v_{\rm cloud}}{2 \mbox{ km/s}}\right) \mu {\rm G},
\label{eq:b1} 
\end{equation} 
where $n_{\rm hot}$ is the number density of the hot plasma in $\mbox{ cm}^{-3}$.
If $v_{\rm cloud} = 30$ km/s, and $n_{\rm hot} = 0.3$,
then $B_{\rm vert}\gg 8 \mu$G. 

As pointed out by Yusef-Zadeh \& Morris (1987a) and Morris \& Serabyn
(1996), if a filament interacts with clouds at more than one point
along its length, as happens in the Radio Arc (Morris 2001), then the
lower limit on the field strength may be considerably greater than in
equation~(\ref{eq:b1}). For certain cloud shapes, a filament could
conceivably avoid the appearance of deformation by rapidly slipping
around the moving clouds.  For some cloud shapes, however, this would
not be possible, and to avoid deformation a filament would have to
dynamically oppose the force applied by the clouds. If this requires
that the magnetic pressure in the filament exceed the ram pressure in
the clouds, then $B_{\rm vert} > 1 $ mG (Yusef-Zadeh \& Morris 1987a).
If a single cloud consists of independently moving clumps, then a
similar argument would apply to a filament interacting with a single
cloud (Morris \& Serabyn 1996).

\section{Conclusion}
\label{sec:discussion}

Because of the simplifying assumptions made in this paper, further
work is needed to determine the extent to which the stability
criterion given by equation~(\ref{eq:stabcrit}) applies to more
realistic equilibria, and to extend the analysis to modes with $k_z
\neq 0$. If equation~(\ref{eq:stabcrit}) does describe, even
approximately, the stability of the boundary between the high and
low-density plasmas in systems with differential rotation, a finite
extent along the $z$ axis, and more general radial profiles---say by
evaluating the terms in equation~(\ref{eq:stabcrit}) near the inner
edge of the dense plasma---then it appears that a pervasive mG
vertical magnetic field can be confined at the Galactic
center. Equation~(\ref{eq:stabcrit}) would then also be relevant to
other accreting systems with strong central poloidal magnetic fields.

A pervasive mG vertical magnetic field in the inner 150 pc of the
Galaxy would have important implications for the origin of the
Galactic magnetic field. If the vertical field originated from radial
inflow over the lifetime of the Galaxy, then the strength of the
average vertical magnetic field in the Galactic disk at the time of
Galaxy formation, $B_i$, can be estimated by magnetic-flux
conservation. If the average mass accretion rate into the
Galactic-center region is 0.3 $M_\sun$/yr (Morris \& Serabyn 1996),
and if the surface density of gas in the disk has remained constant
over the lifetime of the Galaxy, then all the gas that was within
roughly 10 kpc of the Galactic center at the time of Galaxy formation
has been accreted into the Galactic-center region. If the magnetic
field was frozen in to the interstellar plasma, then the vertical
magnetic flux  that initially threaded the central 10 kpc of the disk,
$\Phi_i = B_i \pi (10 \mbox{ kpc})^2$, is now concentrated near
the Galactic center. If there is a pervasive
vertical magnetic field at the Galactic center today, it is unlikely
that it contains regions of oppositely directed vertical fields, since
the parallel geometry of field lines would permit rapid destruction of
the oppositely directed fields by turbulent diffusion and
reconnection. The current magnetic flux through the Galactic-center
region corresponding to such a pervasive field, $\Phi_f$, can thus be
estimated by multiplying field strength times area.  If there is a mG
vertical field throughout the central 150 pc of the Galaxy, then
$\Phi_f =\pi (150 \mbox{ pc})^2 \mbox{ mG}$. Setting $\Phi_i = \Phi_f$
yields $B_i \sim 10^{-7}$~G (Chandran et~al 2000), which would be
consistent with a turbulent dynamo operating in the protogalaxy
(Kulsrud et~al.\ 1997).

\acknowledgements{I thank Amitava Bhattacharjee,
Eric Blackman, Steve Cowley, Ken Gayley, Cornelia Lang, and Mark Morris
for valuable discussions. This work was supported by 
the National Science Foundation under grant AST 00-98086 and
by the Department of Energy under grant DE-FG02-01ER54658 and grant
DE-FC02-01ER54651.}

\appendix

\section{Dispersion relation when
constant-$\beta$ azimuthal-field equilibrium extends from
$r=r_1$ to $r= \infty$} 
\label{ap:inf} 

In this case, the boundary conditions are that the Lagrangian
total-pressure perturbation vanishes at $r=r_1$ and that the behavior
as $r\rightarrow \infty$ is such that the kinetic energy of the mode
is finite. As in section~\ref{sec:bc}, the boundary conditions can be
satisfied in two ways.  The first is for both $c_1$ and $c_2$ to be
non-zero. The boundary condition at $r=r_1$ is then given by the first
row of equation~(\ref{eq:meq}), and a finite kinetic energy requires
that the radicals in equations~(\ref{eq:b1b}) and (\ref{eq:b2b}) be
purely imaginary; if they are real or complex, then the $r^{b_1}$ term
causes the kinetic energy to diverge, where $\sqrt{\dots}$ is taken to
be the root with a positive real part.  The resulting dispersion
relation is the same as equation~(\ref{eq:ldisp1}), but with $p_n$
replaced by $[(0.5q - 1)]^2 + N$, where $N$ is any positive real
number.  The second way is to have $u_{1r} \propto r^b$ with $b= (0.5
q -1) - \sqrt{(0.5q -1)^2 - \Gamma}$, which leads to
equation~(\ref{eq:gdisp1}) and corresponds to a mode in which the
Lagrangian total-pressure perturbation vanishes everywhere.

\section{Differential equation for $k_z=0$ normal modes in the low-density
axial-field region}
\label{ap:axialfield} 

The differential equation for $k_z =0$ normal modes in a uniformly rotating plasma
with axial magnetic field [${\bf B} _0 = B_0(r) \hat{z}$] is
\begin{equation}
\frac{d}{dr} \left[
\frac{\rho_0 (a^2 + v_{\rm A}^2) }{D_{\rm ms} r} \frac{d}{dr}
(r u_{1r})\right]
+ r K u_{1r},
\label{eq:diffeqa}
\end{equation}
where
\begin{equation}
D_{\rm ms} = \omega^2 - \frac{ m^2(a^2 + v_{\rm A}^2)}{r^2},
\label{eq:dms} 
\end{equation}
\begin{equation}
K = - \frac{d}{dr}\left[\frac{m\rho_0 M (a^2 + v_{\rm A}^2)}
{\omega^2 r^2 D_{\rm ms}}\right]
+ \frac{\rho_0 }{r} - \frac{\rho_0}{\omega^2}\frac{d}{dr}\left(\frac{g}{r}\right)
- \frac{M^2 \rho_0}{\omega^2 r D_{\rm ms}},
\label{eq:K} 
\end{equation}
and 
\begin{equation}
M = \frac{mg}{r} - 2\Omega \omega.
\label{eq:M} 
\end{equation}
When $g^\ast = 0$ (i.e., $g = -\Omega^2 r$), 
equation~(\ref{eq:diffeqa}) reduces to the constant-$\Omega$ limit
of equation (3a) of Spies (1978).
In the limit $\rho_0 \rightarrow 0$, with $\rho_0$ constant, 
equation~(\ref{eq:diffeqa}) reduces to 
\begin{equation}
\frac{d}{dr} \left[ r \frac{d}{dr} (r u_{1r})\right] + m^2 u_{1r} = 0,
\label{eq:diffeq2a} 
\end{equation}
with solutions
\begin{equation}
u_{1r} = r^l,
\label{eq:u1ra} 
\end{equation}
with
\begin{equation}
l = -1 \pm m.
\label{eq:l} 
\end{equation} 
Writing the Lagrangian pressure perturbation as $\delta p$ and
the Lagrangian perturbation of $B_z$ as $\delta B_z$, one finds that 
\begin{equation}
\delta p + \frac{B_0 \delta B_z}{4\pi}
= \frac{i r B_0^2 \omega (2+ \gamma \beta)}{8\pi m^2 (a^2 + v_{\rm A}^2)}
\left[ \frac{d}{dr} ( r u_{1r}) + \frac{mM u_{1r}}{\omega^2}\right],
\label{eq:lagpressa} 
\end{equation}
which vanishes as $\rho\rightarrow 0$ for fixed $\omega$, where
$\omega$ is set by the dynamics inside the dense plasma.

\section{Approximate factorization of dispersion relation [equation~(\ref{eq:gdisp1})]
for $|m|=1$.}
\label{ap:m1} 

For the $|m|=1$ roots of equation~(\ref{eq:gdisp1}), the dispersion
relation can be written as 
\begin{equation}
D_{\rm e}\tilde{D}_{\rm R} + S = 0,
\end{equation} 
where $\tilde{D}_{\rm R} = (\omega- \omega_+)(\omega-\omega_-)$, and
\begin{equation}
\omega_\pm = \frac{1}{2\Omega} \left\{
\frac{g}{r} + \frac{a^2}{r^2}\left[ -1 \pm \left(
1 - \frac{g^2}{4a^2\Omega^2}\right) \right] \right\}.
\label{eq:m=1} 
\end{equation}
The value of $S$ can be recalculated and is different from the
values in equations~(\ref{eq:rl}) and (\ref{eq:rg}). 
As in the case $|m|\neq 1$, when $a \ll v_{\rm A}$ and $v_{\rm A} \ll 
v_{\rm rot}$, $ D_{\rm e}\tilde{D}_{\rm R} + S$ has local
minima at $\omega \simeq \pm \sqrt{G/2}$ at which 
$D_{\rm e}\tilde{D}_{\rm R} + S \simeq -G^2/4$, and
stability again turns upon whether 
$D_{\rm e}\tilde{D}_{\rm R} + S $ becomes non-negative
in the interval $\omega \in (-\sqrt{G/2}, \sqrt{G/2})$.
The value of
$D_{\rm e} \tilde{D}_{\rm R}$ has a local maximum at $\omega =
(\omega_+ + \omega_-)/2$ at which $D_{\rm e} \tilde{D}_{\rm R}$ is
always positive. It can be shown that $S\ll D_{\rm e}\tilde{D}_{\rm
R}$ at $\omega = (\omega_+ + \omega_-)/2$, and thus $D_{\rm
e}\tilde{D}_{\rm R} + S$ is always positive at $\omega = (\omega_+ +
\omega_-)/2$, which guarantees that the four exact roots are always
real.

\section{Properties of the eigenmodes}
\label{ap:eigenmodes} 

This appendix treats the eigenmodes corresponding to
equation~(\ref{eq:gdisp1}), for which the Lagrangian total-pressure
perturbation vanishes everywhere.  All of the results of this section
assume that 
\begin{equation}
a\ll v_{\rm A}\ll v_{\rm rot},
\end{equation}
and a number of the
equations to be presented assume that the plasma is not very near
marginal stability, i.e.,
\begin{equation}
\left| \frac{g^2-4\Omega^2 a^2}{g^2+4\Omega^2 a^2} \right| \gg \frac{a}{v_{\rm A}},
\hspace{0.3cm} \mbox{ and } \hspace{0.3cm} 
\left| \frac{g^2-4\Omega^2 a^2}{g^2+4\Omega^2 a^2} \right| \gg \frac{v_{\rm A}}{v_{\rm rot}}.
\label{eq:nnms} 
\end{equation} 

Upon substituting equation~(\ref{eq:gs}) into equation~(\ref{eq:ds}),
one finds that
\begin{equation}
 D_{\rm s} \simeq -\frac{m^2}{r^2 G^2} \left[
2\Omega \sqrt{a^2 G - g^2} \mp \frac{gv_{\rm A}\sqrt{\zeta}}{r}\right]^2,
\label{eq:ds2} 
\end{equation} 
where the $\simeq$ sign has the same meaning as in
equation~(\ref{eq:gs}), that the error in the equation vanishes as
$a/v_{\rm A} \rightarrow 0$ and $v_{\rm A}/v_{\rm rot} \rightarrow
0$. When $\omega$ is real, $(a^2 G- g^2) >0$, $D_{\rm s} $ is real,
and $D_{\rm s} <0$. This implies that all stable modes on the
Rayleigh-Taylor branch of the dispersion relation [equation~(\ref{eq:gs})]
travel more slowly than sound waves. 
Upon substituting equation~(\ref{eq:gs}) into
equation~(\ref{eq:h}), one finds that
\begin{equation}
h \simeq \pm \frac{mv_{\rm A} \sqrt{\zeta}}{r^2 G}
\left( 2\Omega \sqrt{a^2 G - g^2} \mp \frac{gv_{\rm A} \sqrt{\zeta}}{r}
\right) + \frac{ma^2}{r^2}.
\label{eq:h2} 
\end{equation} 
Given equation~(\ref{eq:nnms}), the $ma^2/r^2$ term on the right-hand side
is negligible compared to the other term for $|m| > 1$.
It can be shown that 
\begin{equation}
\left|\frac{h}{D_{\rm s}}\right| \gg 1
\label{eq:hds} 
\end{equation} 
for $|m| > 1$.

When 
\begin{equation}
1 < m \ll v_{\rm rot}/a
\label{eq:mcond}
\end{equation} 
and equation~(\ref{eq:nnms}) is satisfied,  it can be shown that
$|\omega^2 a^2| \ll | D_{\rm s} v_{\rm A}^2|$,
\begin{equation}
b \simeq  - \frac{a^2 m h}{v_{\rm A}^2 D_{\rm s}}
\label{eq:b2} ,
\end{equation} 
and, for unstable modes with $a^2 G - g^2 <0$,
\begin{equation} 
|b| \simeq \frac{a^2}{v_{\rm A}} \sqrt{
\frac{\zeta G}{g^2 - 4\Omega^2 a^2}}.
\label{eq:b3} 
\end{equation} 
The value of $|b|$ in equation~(\ref{eq:b3}) 
is $\ll 1$ when 
$|m|\sim {\cal O}(1)$ and increases
monotonically with increasing $\zeta$, reaching a maximum value of
\begin{equation}
|b|_{\rm max} \simeq \frac{gr}{v_{\rm A}^2} = \alpha
\label{eq:bmax} 
\end{equation} 
at $\zeta _{\rm crit}$,
the largest value of $\zeta$ for which modes are still unstable
[equation~(\ref{eq:zetacrit})].
When $m\sim {\cal O}(1)$, $|b| \ll 1$ for stable modes as well.

Given equations~(\ref{eq:nnms}) and (\ref{eq:mcond}),
it can be shown that 
$|(ma^2/r) du_{1r}/dr| \ll |h u_{1r}|$ and
\begin{equation}
u_{1\theta} \simeq -\frac{ihu_{1r}}{D_{\rm s}}.
\label{eq:u1t2} 
\end{equation} 
Equation~(\ref{eq:hds}) then implies that 
\begin{equation} 
|u_{1\theta}| \gg |u_{1r}|.
\label{eq:u1tu1r} 
\end{equation} 
Equations~(\ref{eq:b2}) and (\ref{eq:u1t2}) can be used to show that
\begin{equation}
\nabla \cdot {\bf u}_1 \simeq \frac{imu_{1\theta}}{r}.
\label{eq:divu} 
\end{equation} 
Since $|\nabla \cdot {\bf u}_1| \gg |u_{1r}/r|$,
the relation (Frieman \& Rotenberg 1960)  
\begin{equation}
p_1 = -\gamma p_0 \nabla \cdot {\bf \xi} - {\bf \xi}\cdot
\nabla p_0
\label{eq:p1fr}
\end{equation}
can be used to show that
\begin{equation}
p_1 \simeq -\gamma p_0 \nabla \cdot {\bf \xi} \simeq
-\frac{ia^2 \rho_0 mh u_{1r}}{\omega r D_{\rm s}}.
\label{eq:p12} 
\end{equation} 
Similarly, the relation
$\rho_1 = -\rho_0 \nabla \cdot {\bf \xi} - {\bf \xi}\cdot \nabla \rho_0$
(Frieman \& Rotenberg 1960) can be used to show that
\begin{equation}
\rho_1 \simeq \frac{\rho_0 m u_{1\theta}}{\omega r} \simeq
-\frac{i\rho_0 mh u_{1r}}{\omega r D_{\rm s}}.
\label{eq:rho12} 
\end{equation} 

If one takes the dot product of equation~(\ref{eq:mom})  with ${\bf B}/B_0$,
uses equation~(\ref{eq:ind}) to eliminate ${\bf B}_1$ and
equation~(\ref{eq:p1fr}) to eliminate $p_1$, one finds that
\begin{equation}
-i\omega \rho_0 u_{1\theta} = -\frac{u_{1r}\rho_0}{\omega} \left(2\Omega \omega - \frac {mg}{r}\right)
- \frac{ma^2\rho_0  \nabla \cdot {\bf u}_1}{\omega r}.
\label{eq:parallel_mom} 
\end{equation} 
The term on the left-hand side represents inertia in the frame of
reference rotating with the plasma, the terms $F_{\rm C,\parallel} =
-u_{1r} 2\Omega \rho_0$, $F_{g,\parallel} = u_{1r} mg\rho_0/\omega r$,
and $F_{p,\parallel} = - ma^2 \rho_0(\nabla \cdot {\bf u}_1)/\omega r$
are the components of the Coriolis, gravitational, and pressure forces
along the perturbed magnetic field line in roughly the $+\hat{\theta}$
direction.  One can show that for modes with complex $\omega$,
\begin{equation}
\left|\frac{F_{p,\parallel}}{F_{g,\parallel}}\right|^2 \simeq 
\frac{a^2 G \zeta}{g^2 \zeta_{\rm crit} }
\label{eq:fpfg} 
\end{equation}
which is less than 1 for $\zeta < \zeta_{\rm crit}$, but which
approaches 1 as $\zeta \rightarrow \zeta_{\rm crit}$ (as $\zeta$ is
increased towards $\zeta_{\rm crit}$, $a^2 G$ is increased towards
$g^2$). The stabilization of modes as $\zeta$ is increased to
$\zeta_{\rm crit}$ thus coincides with pressure becoming equal to
gravity in the dynamics along field lines.  With the use of
equation~(\ref{eq:divu}), equation~(\ref{eq:parallel_mom}) can be
rewritten as
\begin{equation}
-iD_{\rm s} u_{1\theta} \simeq - u_{1r} \left( 2\Omega \omega - \frac{mg}{r}\right).
\label{eq:aparallel_mom} 
\end{equation} 
Equations~(\ref{eq:rho12}) and (\ref{eq:aparallel_mom}) show that the
sign of $D_{\rm s}$ is critical for determining the phase of
$\rho_1$ relative to $u_{1r}$. Let $F \hat{b}$ be the sum of the
components of the gravitational and Coriolis forces along the magnetic
field. As explained in section~\ref{sec:stability},
if $\omega$ is real and $D_{\rm s}<0$, the peaks in the perturbed
density will occur at the points of convergence of $F\hat{b}$.
If $\omega$ is real and $D_{\rm s} > 0$,
the peaks in the perturbed density will occur at the points of divergence
of $F\hat{b}$.

\section{References}

Acheson, D. J., and Gibbons, M. P. 1978, J. Fluid Mech., 85, 743

Balbus, S. A., \& Hawley, J. F. 1998, Rev. Mod. Phys., 70, 1

Bally, J., Stark, A., Wilson, R., and Henkel, C. 1988, ApJ, 324, 223

Benford, G. 1988, ApJ, 333, 735

Bernstein, I., Frieman, E., Kruskal, M., \& Kulsrud, R. 1958,
Proc. Roy. Soc. (London), A244, 17

Chandran, B. D. G., Cowley, S. C., \& Morris, M. 2000, ApJ, 528, 723

Chandrasekhar, S. 1961, {\em Hydrodynamic and hydromagnetic stability}
(New York: Dover)

Chudnovsky, E. M., Field, G. B., Spergel, D. N., \& Vilenkin, A. 1986, Phys. Rev. D, 34, 944

Ferriere, K. M., Zimmer, C., \& Blanc M. 1999, J. Geophys. Res., 104, 17335

Friedberg, J. 1987, {\em Ideal Magnetohydrodynamics} (New York: Plenum; p. 425)

Frieman, E., and Rotenberg, M. 1960, Rev. Mod. Phys., 32, 898 

Gilman, P. 1970, ApJ, 162, 1019

Gonzalez, A., and Gratton, J. 1990, Plasm. Phys. Contr. Nuc. Fus.,
32, 3

Gratton, J., Gratton, F., Gonzalez, A. 1988, Plasma Phys. Contr. Nuc. Fus.,
30, 435

Hawley, J., Balbus, S., \& Stone, J. 2001, ApJ, 554L, 49

Heyvaerts, J., Norman, C., \& Pudritz, R. E. 1988, ApJ, 330, 718

Koyama, K., Maeda, Y., Sonobe, T., Takeshima, T., Tanaka, Y., \&
Yamauchi, S. 1996, Publ. Astron. Soc. Japan, 48, 249

Kulsrud, R. M., Cen, R.,  Ostriker, J. P.,  Ryu, D. 1997, ApJ, 480, 481

Lang, C., Anantharamaiah, K., Kassim, N., \& Lazio, T. 1999, 521, L41

Lang, C., Morris, M., \& Echevarria, L. 1999, ApJ, 526, 727

Lesch, H., \& Reich, W. 1992, A\&A, 264, 493

Lubow, S. H., Papaloizou, J. C. B., \& Pringle, J. E. 1994, MNRAS, 267, 235

Morris, M. 1996, in {\em Unsolved Problems of the Milky Way},
eds. L. Blitz and P. Teuben (Netherlands: IAU,  p. 247)

Morris, M. 2001, private communication

Morris, M., and Serabyn, E. 1996, Ann. Rev. Astron. Astrophys., 34, 645

Newcomb, W. 1961, Phys. Fluids, 4, 391

Newcomb, W. 1962, Nucl. Fusion, suppl., part 2, 451

Press, W., Teukolsky, S., Vetterling, W., \&
Flannery, B. 1992, {\em Numerical Recipes in C}
(Cambridge: Cambridge University Press)

Rayleigh, Lord 1883, Proc. Lond. Math. Soc., 14, 170

Reich, W. 1994, in {\em Nuclei of Normal
Galaxies: Lessons from the Galactic Center}, eds. Genzel, R.,
Harris, A. I. (Dordrecht: Kluwer, p. 55)

Rogers, B., Sonnerup, B. 1986, J. Geophys. Res., 91, 8837

Rosso, F., \& Pelletier, G. 1993, A\&A, 270, 416

Ruzmaikin, A., \& Shukurov, M. 1981, Sov. Astron., 25, 553

Shore, S. N., \& LaRosa, T. N. 1999, ApJ, 521, 587

Sofue, Y., \& Fujimoto, M. 1987, Publ. Astron. Soc. Jpn., 39, 843

Spies, G. O. 1978, Phys. Fluids, 21, 580

Talwar, S. P. 1960, J. Fluid Mech., 9, 581

Terquem, C., \& Papaloizou, J. 1996, MNRAS, 279, 767

Tsuboi, M., Inoue, M. Handa, T., Tabara, H., Kato, T. et~al.\ 1986, 
Astron. J., 92, 818

Tsuboi, M., Kawabata, T., Kasuga, T., Handa, T., Kato, T. 1995,
Publ. Astron. Soc. Jpn, 47, 829

Yamauchi, S., Kawada, M., Koyama, K., Kunieda, H., \&  Tawara, Y. 1990,
ApJ, 365, 532

Yusef-Zadeh, F., and  Morris, M. 1987a, Astron. J., 94, 1178

Yusef-Zadeh, F., and  Morris, M. 1987b, ApJ, 322, 721

\end{document}